\documentclass[sigconf]{acmart}
\newfont{\mycrnotice}{ptmr8t at 7pt}
\newfont{\myconfname}{ptmri8t at 7pt}

\usepackage{mathptmx}
\usepackage{caption}
\usepackage{graphicx}
\usepackage{epstopdf}
\usepackage{epsfig,graphicx}
\usepackage{color}
\usepackage{url}
\usepackage{tabularx}
\usepackage{xfrac}
\usepackage{array}
\usepackage{subfigure}
\usepackage{multirow}
\usepackage{enumitem}
\usepackage{amsmath}
\usepackage{algorithm,algorithmic}
\usepackage{hhline}
 \usepackage{tikz}
\usepackage{booktabs}
\usepackage{adjustbox}
\usepackage{hyperref}
\hypersetup{
     colorlinks   = true,
     citecolor    = gray
}

\begin{document}
\newcommand{\tuan}[1]{{\textcolor{blue}{[\emph{\small Tuan: #1}]}}}
\newcommand{\khoi}[1]{{\textcolor{blue}{[\emph{\small Khoi: #1}]}}}

\newcommand\vpar{{\vspace*{1em}}}
\newcommand{\para}[1]{\noindent{\textbf{#1.}}}
\newcommand{\parai}[1]{\noindent{\textit{#1.}}}
\newcommand{\term}[1]{{\it {\small #1}}}

\newtheorem{mydef}{Definition}

\newcommand{\superscript}[1]{\ensuremath{^{\textrm{#1}}}}
\def\sharedaffiliation{\end{tabular}\newline\begin{tabular}{c}}
\def\ls{\superscript{1}}
\def\xls{\superscript{1,}}
\def\dfki{\superscript{2}}
\def\dk{\superscript{3}}




\title{A Comprehensive Low and High-level Feature Analysis for Early Rumor Detection on Twitter}
\author{Tu Nguyen}
\affiliation{L3S Research Center}
\email{tunguyen@l3s.de}
%
\begin{abstract}

Recent work have done a good job in modeling rumors and detecting them over microblog streams. However, the performance of their automatic approaches are not relatively high when looking early in the diffusion. A first intuition is that, at early stage, most of the aggregated rumor features (e.g., propagation features) are not mature and distinctive enough. The objective of rumor debunking in microblogs, however, are to detect these misinformation as early as possible. In this work, we leverage neural models in learning the hidden representations of individual rumor-related tweets at the very beginning of a rumor. Our extensive experiments show that the resulting signal improves our classification performance over time, significantly within the first 10 hours. To deepen the understanding of these low and high-level features in contributing to the model performance over time, we conduct an extensive study on a wide range of high impact rumor features for the 48 hours range. The end model that engages these features are shown to be competitive, reaches over 90\% accuracy and out-performs strong baselines in our carefully cured dataset.

\end{abstract}


\maketitle

%
%


\vspace{-0.2cm}
\section{Introduction and Background}
\label{sec:intro}
Widely spreading rumors can be harmful to the government, markets and society and reduce the usefulness of social media channel such as Twitter by affecting the reliability of their content. 
Therefore, effective method for detecting rumors on Twitter are crucial and rumors should be detected as early as possible before they widely spread. As an example, let us consider the \textsf{Munich shooting} as a recent event. 
Due to the unclear situation at early time, numerous rumors about the event 
did appear and they started to circulate very fast over social media.
The city police had to warn the population to refrain from spreading related news on Twitter as it was getting out of control: \textit{``Rumors are wildfires that are difficult to put out and traditional news sources or official channels, such as police departments, subsequently struggle to communicate verified information to the public, as it gets lost under the flurry of false information.''}~\footnote{\scriptsize{Deutsche Welle: \url{http://bit.ly/2qZuxCN}}}
%

We follow the rumor definition~\cite{qazvinian2011rumor} considering a rumor (or fake news) as a statement whose truth value is unverified or deliberately false. 
A wide variety of features has been used in existing work in rumor detection 
such as~\cite{castillo2011information,gupta2014tweetcred,jin2013epidemiological,liu2015real,madetecting,ma2015detect,mendoza2010twitter,wu2015false,yang2012automatic}. Network-oriented and other aggregating features such 
as propagation pattern have proven to be very effective for this task. 
Unfortunately, the inherently accumulating characteristic of such features, 
which require some time (and twitter traffic) to mature, does not 
make them very apt for early rumor detection. A first semi-automatic approach focussing 
on early rumor detection presented by Zhao et al.~\cite{zhao2015enquiring}, 
thus, exploits rumor signals such as enquiries that might already arise 
at an early stage. Our fully automatic, cascading rumor detection method follows 
the idea on focusing on early rumor signals on text contents; which is the most reliable source before the rumors widely spread. Specifically, we learn a more complex representation of single tweets using Convolutional Neural Networks, that could capture more hidden meaningful signal than only enquiries to debunk rumors. 
~\cite{madetecting} also use RNN for rumor debunking. However, in their work, RNN is used at \textit{event-level}. The classification leverages only the deep data representations of aggregated tweet contents of the whole event, while ignoring exploiting other --in latter stage--effective features such as user-based features and propagation features. Although, tweet contents are merely the only reliable source of clue at early stage, they are also likely to have doubtful perspectives and different stands in this specific moment. In addition, they could relate to rumorous sub-events (see e.g., the \textsf{Munich shooting}). Aggregating all relevant tweets of the event at this point can be of noisy and harm the classification performance. One could think of a sub-event detection mechanism as a solution, however, detecting sub-events at real-time over Twitter stream is a challenging task~\cite{meladianos2015degeneracy}, which  increases latency and complexity. In this work, we address this issue by deep neural modeling only at single tweet level. Our intuition is to leverage the ``wisdom of the crowd'' theory; such that even a certain portion of tweets at a moment (mostly early stage) are weakly predicted (because of these noisy factors), the ensemble of them would attribute to a stronger prediction. 

The effective cascaded model that engages both low and high-level features for rumor classification is proposed in our other work~\cite{DBLP:journals/corr/abs-1709-04402}. The model uses time-series structure of features to capture their temporal dynamics. In this paper, we make the following contributions with respect to rumor detection:

\begin{itemize}
 	\item We investigate how the performance of different types of low and high-level features changes over time (during the spreading of rumors); improving the understanding of feature impact and model design for rumor detection at different points in time.  
 	\item We also compare our system with human experts. Within 25 hours--the average time for human editors to debunk rumors--we achieve 87\% accuracy.
 \end{itemize}

\section{Single Tweet Credibility Model} 

Before presenting our Single Tweet Credibility Model, we will start 
with an overview of our overall rumor detection method. 
The processing pipeline of our clasification approach is shown in Figure \ref{fig:pipeline}. In the first step, relevant tweets for an event are gathered. Subsequently, in the upper part of the pipeline, 
we predict tweet credibilty with our pre-trained credibility model and aggregate the prediction probabilities on single tweets (CreditScore).
In the lower part of the pipeline, we extract features from tweets and combine them with the creditscore to construct the feature vector in a time series structure called Dynamic Series Time Model. These feature vectors are used to train the classifier for rumor vs. (non-rumor) news  classification.

 \begin{figure}[h]
\centering
\includegraphics[width=0.7\columnwidth]{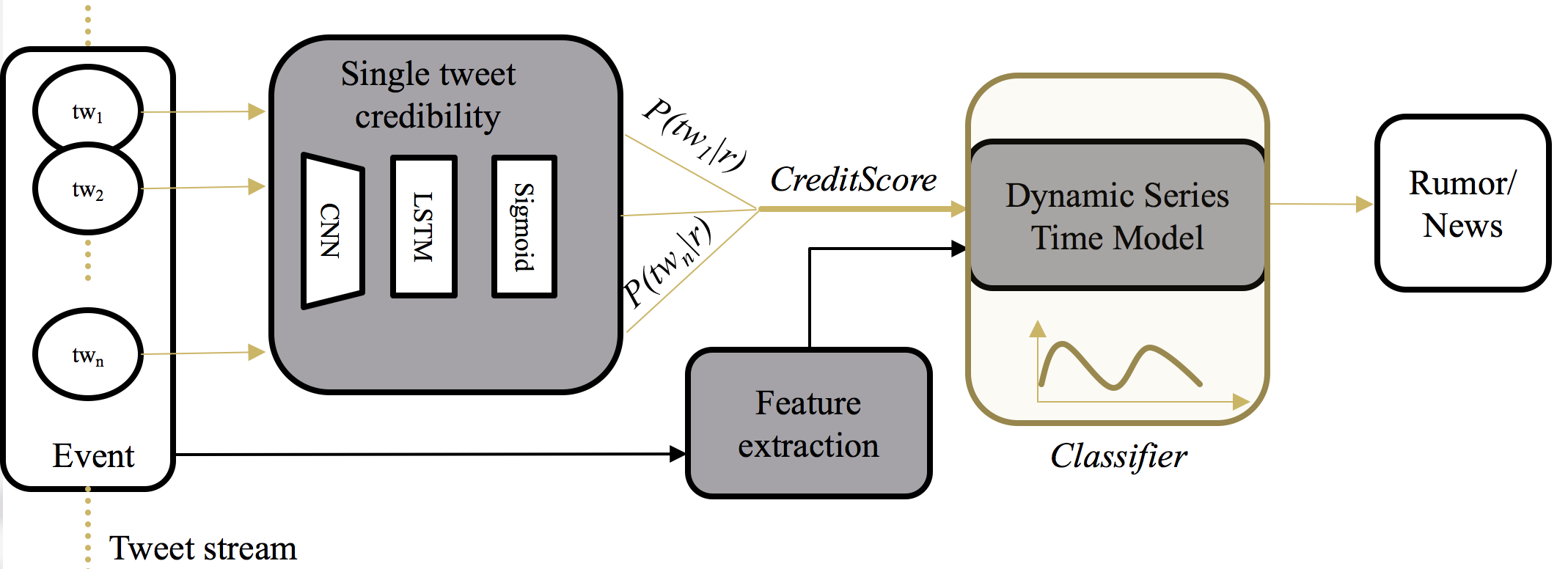}
\caption{Pipeline of our rumor detection approach.}
\label{fig:pipeline}
\end{figure}

Early in an event, the related tweet volume is scanty and there are no clear propagation pattern yet. For the credibility model we, therefore, leverage the signals derived from tweet contents. Related work often uses aggregated content~\cite{liu2015real,ma2015detect,zhao2015enquiring}, since individual tweets are often too short and contain slender context to draw a conclusion. However, content aggregation is problematic for hierarchical events and especially at early stage, in which tweets are likely to convey doubtful and contradictory perspectives. Thus, a mechanism for carefully considering the `vote' for individual tweets is required. In this work, we overcome the restrictions of traditional text representation methods (e.g., bag of words) in handling short text by learning low-dimensional tweet embeddings. In this way, we achieve a rich hidden semantic representation for a more effective classification.

\subsection{Exploiting Convolutional and Recurrent Neural Networks}
\label{subsec:rnn}

Given a tweet, our task is to classify whether it is associated with either a news or rumor. Most of the previous work~\cite{castillo2011information,gupta2014tweetcred} on tweet level only aims to measure the \textit{trustfulness} based on human judgment (note that even if a tweet is trusted, it could anyway relate to a rumor). Our task is, to a point, a reverse engineering task; to measure the probability a tweet refers to a \textit{news} or \textit{rumor} event; which is even trickier. We hence, consider this a weak learning process. Inspired by \cite{zhou2015c}, we combine CNN and RNN into a unified model for tweet representation and classification. The model utilizes CNN to extract a sequence of higher-level phrase representations, which are fed into a long short-term memory (LSTM) RNN to obtain the tweet representation. This model, called CNN+RNN henceforth, is able to capture both local features of phrases (by CNN) as well as global and temporal tweet semantics (by LSTM).

\textbf{Representing Tweets:} 
In this work, we do not use the pre-trained embedding (i.e., \textit{word2vec}), but instead learn the word vectors from scratch from our (large) rumor/news-based tweet collection. The effectiveness of fine-tuning by learning task-specific word vectors is backed by~\cite{kim2014convolutional}. We represent tweets as follows: Let $x_{i} \in \mathcal{R}$ be the $k$-dimensional word vector corresponding to the $i$-th word in the tweet. A tweet of length $n$ (padded where necessary) is represented as: $
x_{1:n} = x_{1} \oplus x_{2} \oplus \cdots \oplus x_{n}
$, where $\oplus$ is the concaternation operator. In general, let $x_{i:i+j}$ refer to the concatenation of words $x_{i},x_{i+1},...,x_{i+j}$. A convolution operation involves a filter $w \in \mathcal{R}^{hk}$, which is applied to a window of $h$ words to produce a feature. For example, a feature $c_{i}$ is generated from a window of words $x_{i:i+h-1}$ by: $ c_{i} = f(w \cdot x_{i:i+h-1} + b)$.

Here $b \in \mathcal{R}$ is a bias term and $f$ is a non-linear function such as the hyperbolic tangent. This filter is applied to each possible window of words in the tweet $\{x_{1:h}, x_{2:h+1},...,x_{n-h+1:n}\}$ to produce a feature map: $c = [c_{1},c_{2},...,c_{n-h+1}]$ with $c \in \mathcal{R}^{n-h+1}$. A max-over-time pooling or dynamic k-max pooling is often applied to feature maps after the convolution to select the most or the k-most important features. We also apply the  1D max pooling operation over the time-step dimension to obtain a fixed-length output.


\textbf{Using Long Short-Term Memory RNNs:} RNN are able to propogate historical information via a chain-like neural network architecture. While processing sequential data, it looks at the current input $x_{t}$ as well as the previous ouput of hidden state $h_{t-1}$ at each time step. However, standard RNNs become unable to learn long-term dependencies, when the gap between two time steps becomes large. To address this issue, LSTM was introduced in~\cite{hochreiter1997long}. The LSTM architecture has a range of repeated modules for each time step as in a standard RNN. At each time step, the output of the module is controlled by a set of gates in $\mathcal{R}^{d}$ as a function of the old hidden state $h_{t-1}$ and the input at the current time step $x_{t}$: forget gate $f_{t}$, input gate $i_{t}$, and output gate $o_{t}$. 
\subsection{CNN+LSTM for tweet-level classification.}
We regard the output of the hidden state at the last step of LSTM as the final tweet representation and we add a softmax layer on top. We train the entire model by minimizing the cross-entropy error. Given a training tweet sample $x^{(i)}$, its true label $y_{j}^{(i)} \in \{y_{rumor},y_{news}\}$ and the estimated probabilities $\tilde{y}_{j}^{(i)} \in [0..1]$ for each label $j \in \{rumor,news\}$, the error is defined as:

\begin{equation}
\mathsf{L}(x^{(i)},y^{(i)}) = 1\{y^{(i)}=y_{rumor}\}log(\tilde{y}_{rumor}^{(i)}) + 1\{y^{(i)}=y_{news}\}log(\tilde{y}_{news}^{(i)})
\end{equation}

where $1$ is a function converts boolean values to $\{0,1\}$. We employ stochastic gradient descent (SGD) to learn the model parameters.

\section{Time Series Rumor Detection Model} 
\label{sec:timr_seriers_rumor_model}
As observed in~\cite{madetecting,ma2015detect}, rumor features are very prone to change during an event's development. In order to capture these temporal variabilities, we build upon the Dynamic Series-Time Structure (DSTS) model (time series for short) for feature vector representation proposed in~\cite{ma2015detect}. We base our features on the time series approach and train the classifier with those features as well as with the credibility model's predictions in a cascaded manner. In this section, we first detail the employed Dynamic Series-Time Structure, then describe the features used for learning in this pipeline step.

  \subsection{ Dynamic Series-Time Structure (DSTS) Model} 

%

For an event $E_i$ we define a time frame given by $timeFirst_i$ as the start time of the event and $timeLast_i$ as the time of the last tweet of the event in the observation time.
We split this event time frame into N intervals and associate each tweet to one of the intervals according to its creation time.  Thus, we can generate a vector V($E_i$) of features for each time interval. In order to capture the changes of feature over time, we model their differences between two time intervals. So the model of DSTS is represented as: $V(E_i)=(\textbf{F}^D_{i,0}, \textbf{F}^D_{i,1},..., \textbf{F}^D_{i,N},\textbf{S}^D_{i,1},..., \textbf{S}^D_{i,N})$, where $\textbf{F}^D_{i,t}$ is the feature vector in time interval t of event $E_i$. $\textbf{S}^D_{i,t}$ is the difference between $\textbf{F}^D_{i,t}$ and $\textbf{F}^D_{i,t+1}$. V($E_i$ ) is the time series feature vector of the event $E_i$.
\begin{equation}
\textbf{F}^D_{i,t}=(\widetilde{ f}_{i,t,1},\widetilde{ f}_{i,t,2},...,\widetilde{ f}_{i,t,D})
\end{equation}
\begin{equation}
\textbf{S}^D_{i,t}=\frac{\textbf{F}^D_{i,t+1}-\textbf{F}^D_{i,t}}{Interval(E_i)}
\end{equation}
We use Z-score to normalize feature values;
$
\widetilde{f}_{i,t,k}=\frac{f_{i,t+1,k}-\overline{f}_{i,k}}{\sigma(f_{i,k})}
$
 where $f_{i,t,k}$ is the k-th feature of the event $E_i$ in time interval t. The mean of the feature k of the event $E_i$ is denoted as  $\overline{f}_{i,k}$ and $\sigma(f_{i,k})$ is the standard deviation of the feature k over all time intervals. We can skip this step, when we use Random Forest or Decision Trees, because they do not require feature normalization.


%

\subsection{Features for the Rumor Detection Model} 
\label{sub:features}

In selecting features for the rumor detection model, we have followed 
two rationales: a) we have selected features that we expect to be useful 
in early rumor detection and b) we have collected a broad 
range of features from related work as a basis for investigating 
the time-dependent impact of a wide varietey of features 
in our time-dependence study. 
In total, we have constructed over 50 features\footnote{details are in Table~\ref{tab:full_features}.}
in the three main categories i.e., \textit{Ensemble}, \textit{Twitter} and \textit{Epidemiological} features.  
We refrained from using network features, since they are  expected to 
be of little use in early rumor detection, since user networks around events 
need time to form. Following our general idea, none of our features are extracted from the content aggregations.
 
 \subsubsection{Ensemble Features.} 
 \label{sub:ef} \indent
 We consider two types of Ensemble Features: features accumulating crowd wisdom and averaging feature for the Tweet credit Scores. 
The former are extracted from the surface level while the latter comes from the low dimensional level of tweet embeddings; that in a way augments the sparse crowd at early stage.

\textbf{\textit{CrowdWisdom.}} Similar to~\cite{liu2015real}, the core idea is to leverage the public's common sense for rumor detection: If there are more people denying or doubting the truth of an event, this event is more likely to be a rumor. For this purpose, ~\cite{liu2015real} use an extensive list of bipolar sentiments with a set of combinational rules. In contrast to mere \textit{sentiment} features, this approach is more tailored \textit{rumor} context (difference not evaluated in~\cite{liu2015real}). We simplified and generalized the ``dictionary'' by keeping only a set of carefully curated \textit{negative words}. We call them ``debunking words" e.g., \textit{hoax}, \textit{rumor} or \textit{not true}. Our intuition is, that the attitude of doubting or denying events is in essence sufficient to distinguish rumors from news. What is more, this generalization augments the size of the crowd (covers more 'voting' tweets), which is crucial, and thus contributes to the quality of the crowd wisdom. In our experiments, ``debunking words" is an high-impact feature, but it needs substantial time to ``warm up"; that is explainable as the crowd is typically sparse at early stage. 

\textbf{\textit{CreditScore.}}
The sets of single-tweet models' predicted probabilities are combined using an \textit{ensemble averaging}-like technique. In specific, our pre-trained $CNN+LSTM$ model predicts the credibility of each tweet $tw_{ij}$ of event $E_{i}$. The \textit{softmax} activation function outputs probabilities from 0 (rumor-related) to 1 (news). Based on this, we calculate the average prediction probabilities of all tweets $tw_{ij} \in E_{i}$ in a time interval $t_{ij}$. In theory there are different sophisticated ensembling approaches for averaging on both training and test samples; but in a real-time system, it is often convenient (while effectiveness is only affected marginally) to cut corners. In this work, we use a sole training model to average over the predictions. We call the outcome \textsf{CreditScore}. 

\subsubsection{Twitter-based Features.}\indent 
\label{sub:tb} - consist of \textit{text}, \textit{Twitter} and \textit{user} features. The features in this group are aggregated feature values extracted from tweet-level of all relevant tweets in a time interval. 
\textbf{\textit{Text Features}} are derived from a tweet's text content. We consider 16 text features including \textsf{lengthOftweet} and \textit{smile} (contain $:->, :-), ;->, ;-)..)$, \textit{sad}, \textit{exclamation}, \textit{I-you-heshe} (contain first, second, third pronouns). In addition, we use the natural language Toolkit (nlTK)\footnote{http://www.nltk.org/} to analyze the tweets' sentiment and extract the following features: the \emph{NumPositiveWords}, \emph{NumNegativeWords} and \emph{PolarityScores}. \emph{PolarityScores} is a float for sentiment strength of one tweet\footnote{http://www.nltk.org/api/nltk.sentiment.html}. $Polarity\_scores = \frac {1}{N}    \sum_{0}^{n} {Polarity(token_n)}$.
%

\textbf{\textit{Twitter Features}} refer to basic Twitter features, such as \textit{hashtags}, \textit{mentions}, \textit{retweets}. In addition, we derive three more URL-based features. The first is the WOT--trustworthy-based-- score which is crawled from the APIs of WOT.com\footnote{https://www.mywot.com/en/api}. The second is domain categories which we have collected from BLUECOAT.com\footnote{http://sitereview.bluecoat.com/} and grouped them into 2 types: \textit{news} and \textit{non-news}. The last last one is based on the rank of the domain retrieved from ALEXA.com\footnote{http://www.alexa.com/siteinfo/bbc.com}. Here we also use a split into 2 groups: domains ranked lower than 5000 and those ranked higher. 
Figure~\ref{fig:url5000} illustrates the fractions of URLs in tweets with rank lower than 5000 over time, for news and rumors. It could be understood that rumors are more likely spreaded by less reliable sources.
  \begin{figure}[h]
\centering
\includegraphics[width=0.6\columnwidth]{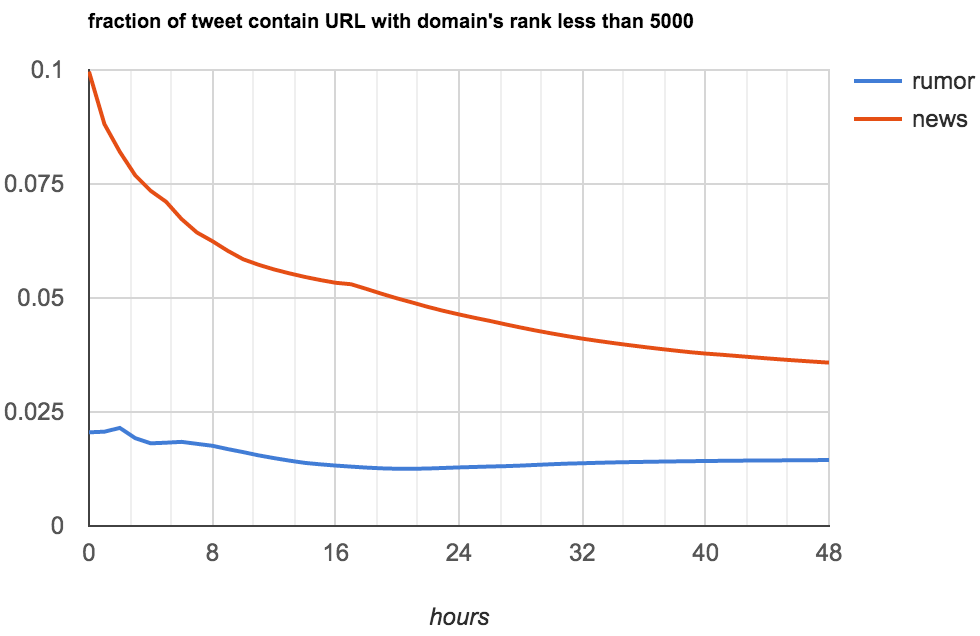}

\caption{The fraction of tweets contain URLs with domain's rank less than 5000.}
\label{fig:url5000}
\end{figure}

\textbf{\textit{User Features.}} Apart from the features already exploited in related work (e.g., \textit{VerifiedUser}, \textit{NumOfFriends}, \textit{NumOfTweets}, \textit{ReputationScore}), we add two new features captured from Twitter interface:  (1) how many photos have been posted by a user \emph{(UserNumPhoto)}, and (2) whether the user lives in a large city. We use the list of large cities in a report by demographia\footnote{http://www.demographia.com/db-worldua.pdf}. Our empirical results on this interesting feature indicate a larger margin of news posters live in large cities (17,5\%) compared to rumor posters (13\%). The fraction of rumor posters stays high (comparable with news) at the very first hour, but then falling dramatically.

\subsubsection{Epidemiological Modeling Features}
\label{sec:epide}
he  the Epidemiological features are based on propagation models adopted from epidemiology, which are widely used for conventional rumor detection. These models have been shown to work well for long running events, yet we expect a lower performance for early detection. In~\cite{jin2013epidemiological}, the propagation pattern of rumors and news is modeled by the adaptation of two propagation models: SIS (Susceptible, Infected, Susceptible) and SEIZ (susceptible, exposed, infected, skeptic). 
To adapt to the scenario of Twitter, a user who posts a tweet of relevant event is defined as \textbf{(I)} infected, a user who does not is defined as \textbf{(S)} susceptible. But unlike the normal epidemiological scenario where infected nodes can be cured and return to susceptible, in Twitter the user once posts a event-related tweet, will be classified into the infected component and cannot return to susceptible again. At time $t$, the total number of population is $\Delta N(t)= I(t) + S(t)$ where $I(t)$ is the size of infected population and $S(t)$ is the size of susceptible population. SIS model assumes that a susceptible user once exposed to a infected user turns to infected immediately. However, when Twitter users see tweets, they have their own judgment on the credibility and they can decide whether further spreading the tweet or ignoring them.

 To adapt to Twitter context, the compartments of the SEIZ model~\cite{bettencourt2006power} can be mapped like this: \textbf{(S)}usceptible is a user who has not been exposed to the event, in the other word, he does not see any tweets about the certain event yet. \textbf{(I)}nfected  means a user has posted event-related tweets, \textbf{(Z)} skeptic is a user who has been exposed to the certain event but he decides to ignore it and \textbf{(E)}xposed is a user who has been exposed to the certain event but he will post the tweets after some delay.

We use Levenberg-Marquardt algorithm to learn the parameters of the SIS and SEIZ. In each time interval from $t_0$ to $t_n$, we fit the sequenced  tweets' volume from the beginning time $t_0$ to the current time interval $t_n$ of an event to SIS and SEIZ model and learn the parameters. From SIS we get two features $\beta_{n},\alpha_{n}$ and from SEIZ we get 7 features $\beta_{n},b_{n},l_{n},p_{n},\varepsilon_{n},\rho_{n},RSI_{n}$. Look at Table~\ref{tab:full_features} for details. 
 We show 2 examples as following two rumors and news in Figures \ref{fig:SISModel}  
 It is obvious that SEIZ is more appropriate than SIS to model in our application of Twitter, because the fitting error of SEIZ is less than SIS.

%
%

\begin{figure}[!h]

  \centering

\subfigure[SIS and SEIZ model for rumor 1]{\label{fig:SIS-rumor1}
\centering
  \includegraphics[width=0.45\columnwidth]{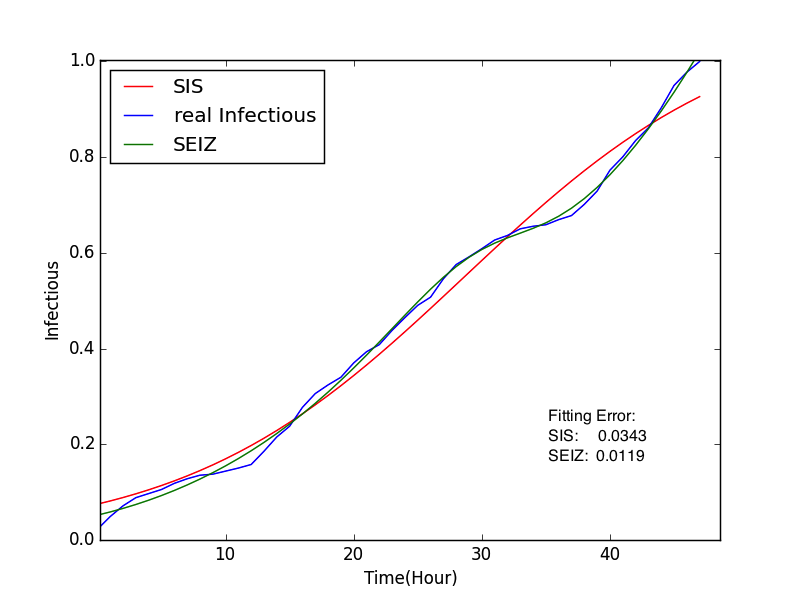}
} %
\subfigure[SIS and SEIZ model for news 1]{\label{fig:SIS-news1}
\centering
  \includegraphics[width=0.45\columnwidth]{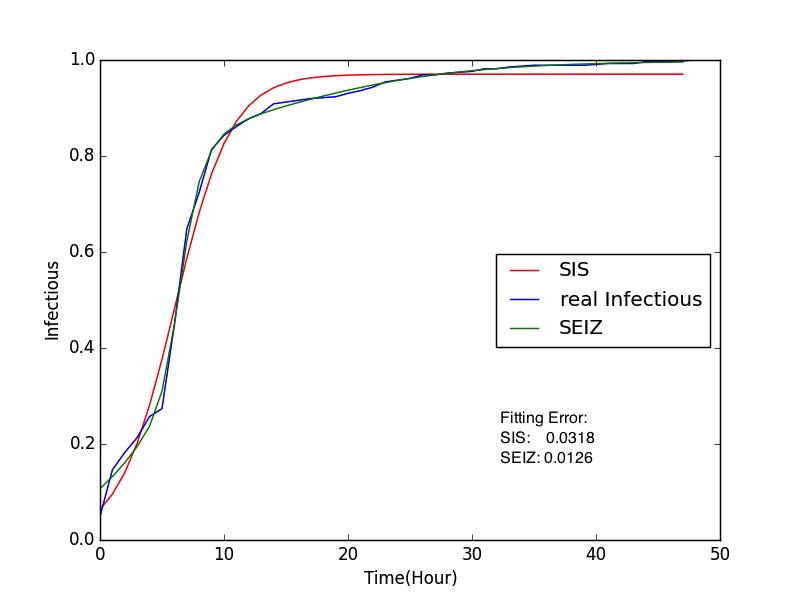}
}
\caption{Fitting results of SIS and SEIZ model of (a) Rumor: Robert Byrd was a member of KKK 
(b) News: Doctor announces Michael Schumacher is making progress.}
\label{fig:SISModel}
\end{figure}

But if we fit the models of the first few hours with limited data, the result of learning parameters is not so accurate.  We show the performance of fitting these two model with only the first 10 hours tweets' volume in Figure \ref{fig:SISModelshort}. As we can see except for the first one, the fitting results of other three are not good enough.

\begin{figure}[!h]

  \centering

\subfigure[SIS and SEIZ Model for Rumor \#1 with 10 Hours Data]{\label{fig:SIS-rumor1}
\centering
  \includegraphics[width=0.45\columnwidth]{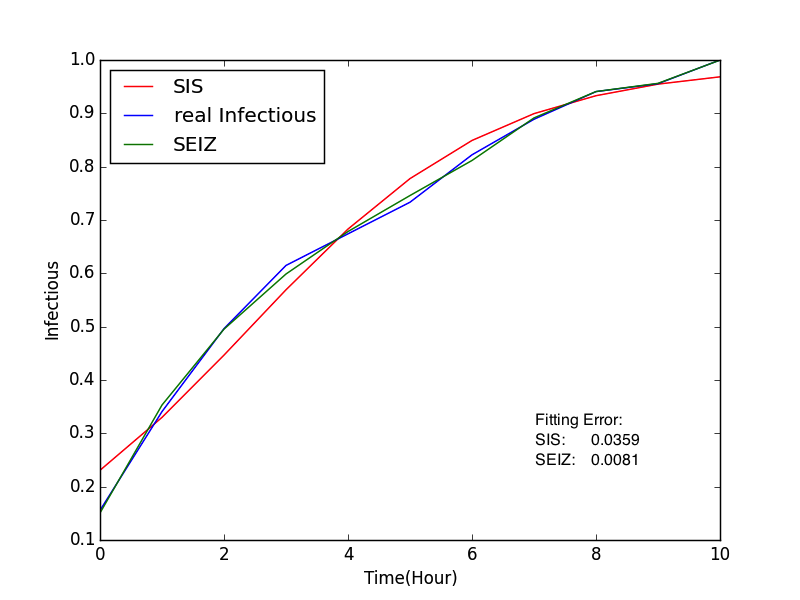}
} %
\subfigure[SIS and SEIZ Model for News \#1 with 10 Hours Data]{\label{fig:SIS-news1}
\centering
  \includegraphics[width=0.45\columnwidth]{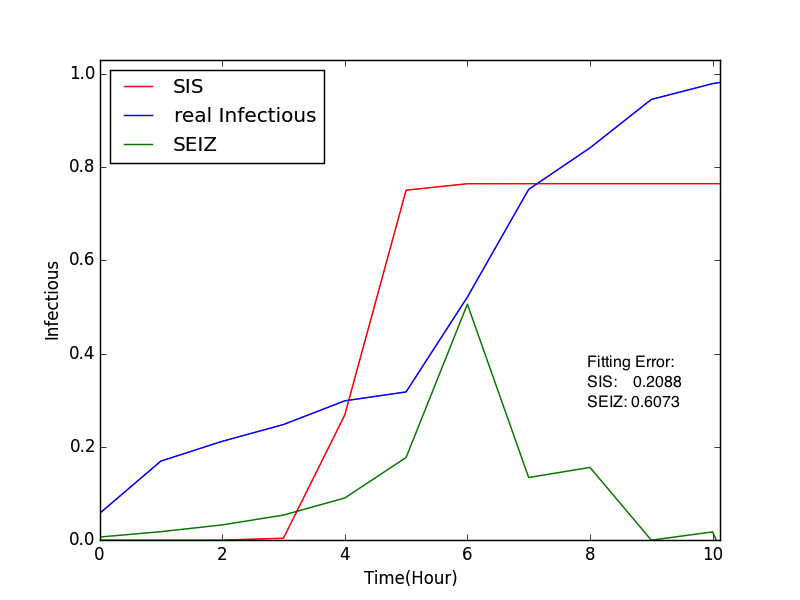}
}
\caption{Fitting Results of SIS and SEIZ Model with Only First 10 Hours Tweets' Volume (same 2 stories as above)}
\label{fig:SISModelshort}
\end{figure}

\textbf{SpikeM model Features} Kwon et al.~\cite{kwon2013prominent} discuss a further approach that extend SpikeM~\cite{conf/kdd/MatsubaraSPLF12}, so that it can describe multi-pike pattern of information diffusion. SpikeM extends the SI model from: 

%
%

 \begin{itemize}
\item a power-law decay term $f(\tau)=\beta \cdot \tau ^{-1.5}$ 
So the infection strength of the nodes which are earlier infected decreases in with a power-law decay pattern.

\item a periodic interaction function.
It stands for that people have a periodic interaction patterns, for example people go to sleep at night, so they post much less tweets at night. Parameters $P_p$, $P_a$, and $P_s$ are the period, strength, and shift of the periodic interaction function.
\item $\varepsilon$ is the background noise term. 
\end{itemize}
But the SpikeM can't fit to the events with multi-pikes. For that, the term external shock $S(n)$ should not occur once but more. So~\cite{kwon2013prominent} extend the SpikeM by adding a periodic interaction function for the term external shock $S(n)$. Same approach as fitting SIS model, we learn the parameters of SpikeM model with Levenberg-Marquardt algorithm. We show 2 examples of the adapted SpikeM fitting result in Figure \ref{fig:SPikeModel}. But SpikeM has the same problem as fitting SIS or SEIZ model, if we test only within 10 hours data the results are much worse than the results with full 48 hours showing in Figure \ref{fig:SpikeMModelshort}.

\begin{figure}[!h]

  \centering

\subfigure[SpikeM Model for Rumors \#1]{\label{fig:SPikeM-rumor1}
\centering
  \includegraphics[width=0.45\columnwidth]{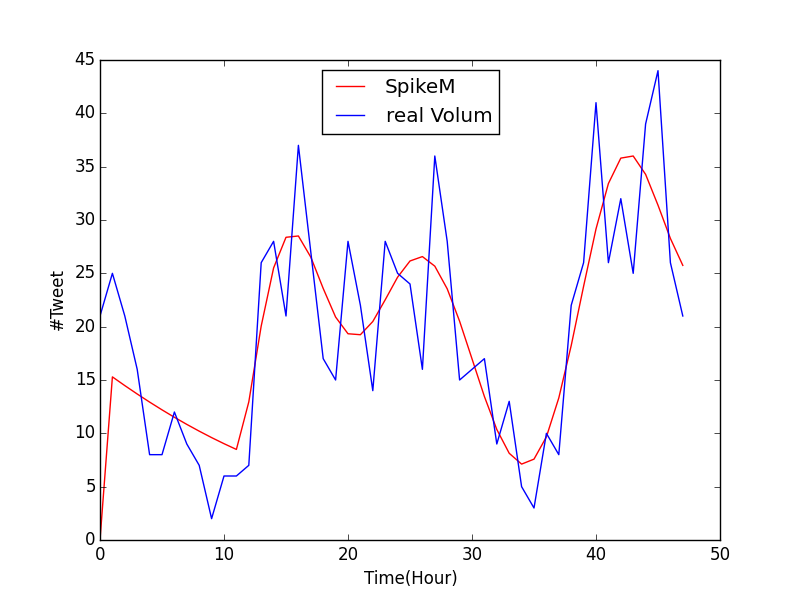}
} %
\subfigure[SpikeM Model for News \#1]{\label{fig:SPikeM-news1}
   \includegraphics[width=0.45\columnwidth]{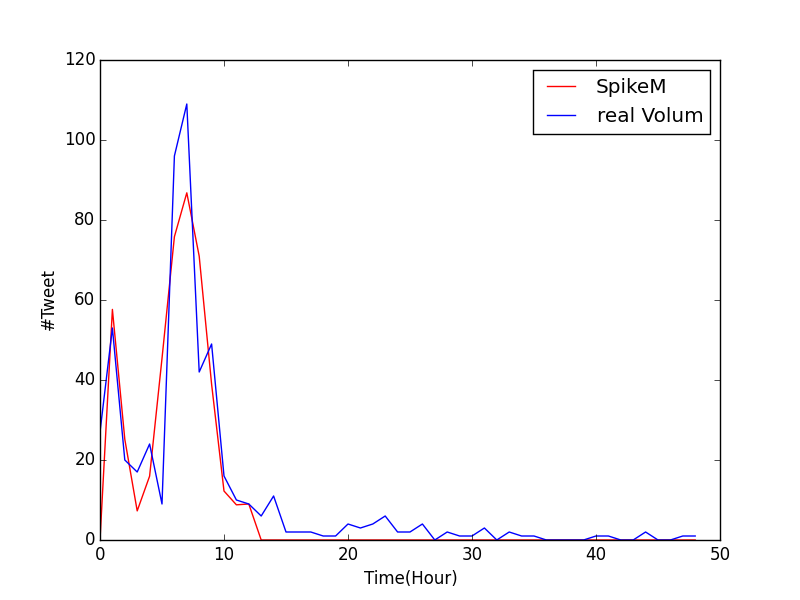}
}
%
\caption{Fitting Results of SpikeM Model of (a) Rumor: Robert Byrd was a member of KKK and (b) News: Doctor announces Michael Schumacher is making process}
\label{fig:SPikeModel}
\end{figure}

\begin{figure}[!h]

  \centering

\subfigure[SpikeM Model for Rumor \#1 with 10 Hours Data]{\label{fig:SPikeM-rumor1}
\centering
  \includegraphics[width=0.45\columnwidth]{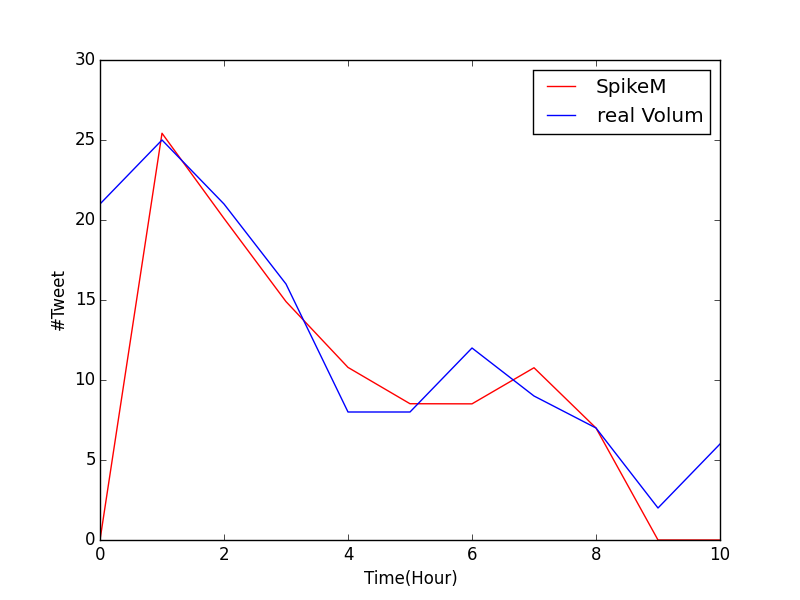}
} %
\subfigure[SpikeM Model for News \#1 with 10 Hours Data]{\label{fig:SPikeM-news1}
\centering
  \includegraphics[width=0.45\columnwidth]{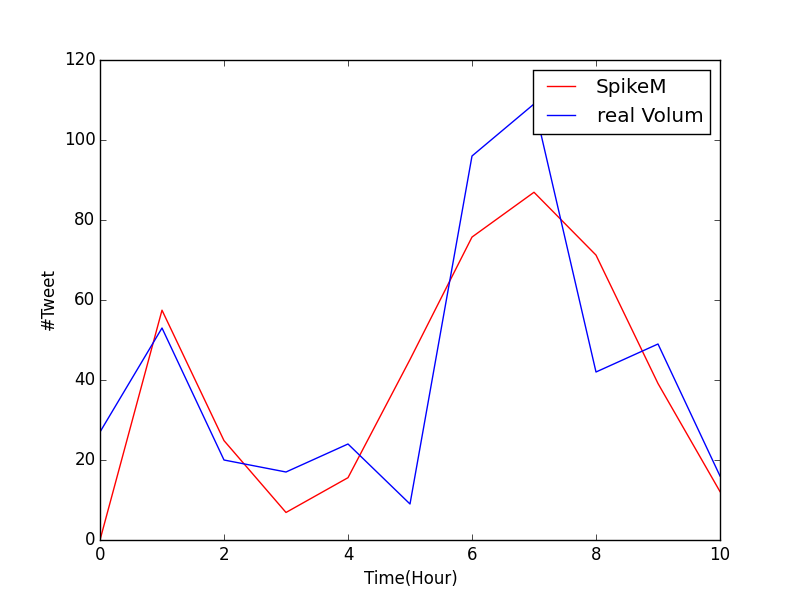}
}

\caption{Fitting results of SpikeM Model with First 10 Hours data (same stories as above) }
\label{fig:SpikeMModelshort}
\end{figure}

\begin{table}[!h]
\small
\centering
\scalebox{0.6}{
 \begin{tabular}{@{}lllllll@{}}
 \toprule
 \textbf{Category} & \textbf{Feature} & \textbf{Description}\\ \midrule
 Twitter & Hashtag & \% tweets contain \#hashtag  \cite{castillo2011information}\cite{liu2015real}\cite{qazvinian2011rumor}\cite{gupta2014tweetcred}\cite{liu2015real}\\
 	Features	& Mention &  \% tweets mention others @user  \cite{castillo2011information}\cite{liu2015real}\cite{qazvinian2011rumor}\cite{gupta2014tweetcred}\cite{liu2015real}\\
 		& NumUrls &  \# URLs in the tweet  \cite{castillo2011information}\cite{qazvinian2011rumor}\cite{gupta2014tweetcred}\cite{yang2012automatic}\cite{liu2015real}\\
 		& Retweets & average \# retweets \cite{liu2015real} \\ 
 		& IsRetweet & \% tweets are retweeted from others \cite{castillo2011information}\cite{gupta2014tweetcred}\\
 		& ContainNEWS & \% tweets contain URL and its domain's catalogue is News \cite{liu2015real}\\
 		& WotScore & average WOT score of domain in URL \cite{gupta2014tweetcred}\\
 		& URLRank5000 & \%  tweets contain URL whose domain's rank less than 5000 \cite{castillo2011information}\\
 		& ContainNewsURL & \% tweets contain URL whose domain is News Website\\
\midrule
 Text  & LengthofTweet & average tweet lengths \cite{castillo2011information}\cite{gupta2014tweetcred}\\
 Features   & NumOfChar & average \# tweet characters \cite{castillo2011information}\cite{gupta2014tweetcred}\\
   & Capital &  average fraction of characters in Uppercase \cite{castillo2011information} \\
   & Smile & \% tweets contain :->, :-), ;->, ;-) \cite{castillo2011information}\cite{gupta2014tweetcred}\\
   & Sad & \%  tweets contain :-<, :-(, ;->, ;-( \cite{castillo2011information}\cite{gupta2014tweetcred}\\
   & NumPositiveWords & average \# positive words \cite{castillo2011information}\cite{gupta2014tweetcred}\cite{yang2012automatic}\cite{liu2015real}\\
   & NumNegativeWords & average \# negative words \cite{castillo2011information}\cite{gupta2014tweetcred}\cite{yang2012automatic}\cite{liu2015real}\\
   & PolarityScores & average polarity scores of the Tweets \cite{castillo2011information}\cite{yang2012automatic}\cite{liu2015real}\\
   & Via & \% of tweets contain via \cite{gupta2014tweetcred}\\
   & Stock & \% of tweets contain \$  \cite{castillo2011information}\cite{gupta2014tweetcred}\\
   & Question & \% of tweets contain ? \cite{castillo2011information}\cite{liu2015real}\\
   & Exclamation & \% of tweets contain ! \cite{castillo2011information}\cite{liu2015real}\\
   & QuestionExclamation & \% of tweets contain multi Question or Exclamation mark \cite{castillo2011information}\cite{liu2015real}\\ 
   & I & \% of tweets contain first pronoun like I, my, mine, we, our    \cite{castillo2011information}\cite{gupta2014tweetcred}\cite{liu2015real}\\
   & You & \% of tweets contain second pronoun like U, you, your, yours  \cite{castillo2011information}\\ 
   & HeShe & \% of tweets contain third pronoun like he, she, they, his, etc.  \cite{castillo2011information}\\ \midrule
   User & UserNumFollowers  & average number of followers \cite{castillo2011information}\cite{gupta2014tweetcred}\cite{liu2015real}\\
  Features	& UserNumFriends  & average number of friends \cite{castillo2011information}\cite{gupta2014tweetcred}\cite{liu2015real}\\
 	& UserNumTweets  & average number of users posted tweets \cite{castillo2011information}\cite{gupta2014tweetcred}\cite{yang2012automatic}\cite{liu2015real}
\\
 	& UserNumPhotos  & average number of users posted photos \cite{yang2012automatic}\\
 	& UserIsInLargeCity  & \% of users living in large city \cite{yang2012automatic}\cite{liu2015real}\\
 	& UserJoinDate & average days since users joining Twitter \cite{castillo2011information}\cite{yang2012automatic}\cite{liu2015real}
\\
 	& UserDescription  & \% of user having description \cite{castillo2011information}\cite{yang2012automatic}\cite{liu2015real}
\\
 	& UserVerified  & \% of user being a verified user\cite{yang2012automatic}\cite{liu2015real}
\\
 	& UserReputationScore & average ratio of \#Friends over (\#Followers + \#Friends) \cite{liu2015real}\\   \midrule
 Epidemiological & $\beta_{SIS}$ & Parameter $\beta$ of Model SIS \cite{jin2013epidemiological}\\
 				 Features			& $\alpha_{SIS} $ & Parameter $\alpha$ of Model SIS \cite{jin2013epidemiological}\\
 							& $\beta_{SEIZ}$ & Parameter $\beta$ of Model SEIZ \cite{jin2013epidemiological}\\
 							& $b_{SEIZ}$ & Parameter b of Model SEIZ\cite{jin2013epidemiological}\\
 							& $l_{SEIZ}$ & Parameter l of Model SEIZ \cite{jin2013epidemiological}\\
 							& $p_{SEIZ}$ & Parameter p of Model SEIZ \cite{jin2013epidemiological}\\
 							& $\varepsilon_{SEIZ}$ & Parameter $\varepsilon$ of Model SEIZ \cite{jin2013epidemiological}\\
 							& $\rho_{SEIZ}$ & Parameter $\rho$ of Model SEIZ \cite{jin2013epidemiological}\\
 							& $R_{SI}$ & Parameter $R_{SI}$ of Model SEIZ \cite{jin2013epidemiological}\\
		\midrule	
 SpikeM & $P_s$ & Parameter $P_s$ of Model Spike \cite{kwon2013prominent}\\
 			 Model 				& $P_a$ & Parameter $P_a$ of Model SpikeM \cite{kwon2013prominent}\\
 					Features		& $P_p$ & Parameter $P_p$ of Model SpikeM \cite{kwon2013prominent}\\
 							& $Q_s$  & Parameter $Q_s$ of Model SpikeM \cite{kwon2013prominent}\\
 							& $Q_a$ & Parameter $Q_a$ of Model SpikeM \cite{kwon2013prominent}\\
 							& $Q_p$ & Parameter $Q_p$ of Model SpikeM \cite{kwon2013prominent}\\ \midrule	
 Crowd Wisdom  & CrowdWisdom & \% of tweets containing "Debunking Words" \cite{liu2015real} \cite{zhao2015enquiring}\\ \midrule
 CreditScore  & CreditScore & average CreditScore\\
 \bottomrule
 \end{tabular}}
 \caption{Features of Time Series Rumor Detection Model}
 \label{tab:full_features}
\end{table}

\section{Experimental Evaluation} 
  \subsection{Data Collection} 
  \label{sec:dataset_single}
To construct the training dataset, we collected rumor stories from the rumor tracking websites \textbf{snopes.com} and \textbf{urbanlegends.about.com}. In more detail, we crawled 4300 stories from these websites. From the 
story descriptions we manually constructed queries to retrieve the relevant tweets for the 270 rumors with highest impact. Our approach to query construction mainly follows~\cite{gupta2014tweetcred}. For the news event instances (non-rumor examples), we make use of the corpus from Mcminn et al.~\cite{mcminn2013building}, which covers 500 real-world events. They have crawled tweets via the streaming API from $10^{th}$ of October 2012 to $7^{th}$ of November 2012. The involved events have been manually verified and related to tweets with relevance judgments, which has resulted in a high quality corpus. From the 500 events, we select top 230 events with the highest tweet volumes (as a criteria for event impact). Furthermore, we have added 40 other news events, which happened around the time periods of our rumors. This results in a dataset of 270 rumors and 270 events.  The dataset details are shown in Table \ref{tab:Tweet_Volume}. We then constructs two distinct datasets for (1) single tweet and (2) rumor classification. 

\begin{table}[t]
 \centering
\scalebox{0.8}{
 \begin{tabular}{@{}cccccc@{}}
 \toprule
 \textbf{Type} & \textbf{Min Volume} & \textbf{Max Volume}& \textbf{Total} &\textbf{Average} \\ \midrule
 News & 98 & 17414 & 345235 & 1327.82 \\ \midrule
 Rumors & 44  & 26010& 182563 & 702.06\\
 	 \bottomrule
 \end{tabular}}
 \caption{Tweet Volume of News and Rumors}
 \label{tab:Tweet_Volume}
\end{table}
 \textbf{Training data for single tweet classification.}  An event might include sub-events for which relevant tweets are rumorous. To deal with this complexity, we train our single-tweet learning model only with manually selected \textit{breaking and subless} events from the above dataset. In the end, we used 90 rumors and 90 news associated with 72452 tweets, in total. This results in a highly-reliable ground-truth of tweets labelled as \textit{news}-related and \textit{rumor}-related, respectively. Note that the labeling of a tweet is inherited from the event label, thus can be considered as an semi-automatic process.

\subsubsection{Time Period of an Event}
\label{sec:Time_Period_of_an_Event}
The time period of a rumor event is sometimes fuzzy and hard to define. One reason is a rumor may have been triggered for a long time and kept existing, but it did not attract public attention. However it can be triggered by other events after a uncertain time and suddenly spreads as a bursty event. E.g., a rumor\footnote{http://www.snopes.com/robert-byrd-kkk-photo/} claimed that Robert Byrd was member of KKK. This rumor has been circulating in Twitter for a while. As shown in Figure \ref{fig:KKK_full} that almost every day there were several tweets talking about this rumor. But this rumor was triggered by a picture about Robert Byrd kissing Hillary Clinton in 2016~\footnote{http://www.snopes.com/clinton-byrd-photo-klan/} and Twitter users suddenly noticed this rumor and it was spreaded burstily. In this work, what we are really interested in is the tweets which are posted in hours around the bursty peak. We defined the hour with the most tweets' volume as $t_{max}$ and we want to detect the rumor event as soon as possible before its burst, so we define the time of the first tweet before $t_{max}$ within 48 hours as the beginning of this rumor event, marked as $t_{0}$. And the end time of the event is defined as $t_{end}=t_0+48$. We show the tweet volumes in Figure \ref{fig:KKK_part} of the above rumor example.

\begin{figure}[!h]
\centering
\subfigure[Before]{\label{fig:KKK_full}
\centering
\includegraphics[width=0.4\columnwidth]{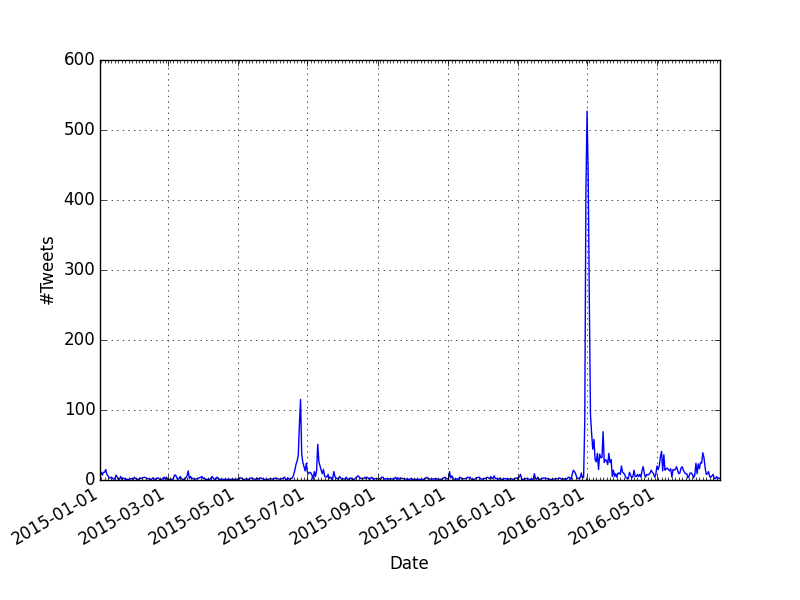}
}
\subfigure[After]{\label{fig:KKK_part}
\centering
\includegraphics[width=0.4\columnwidth]{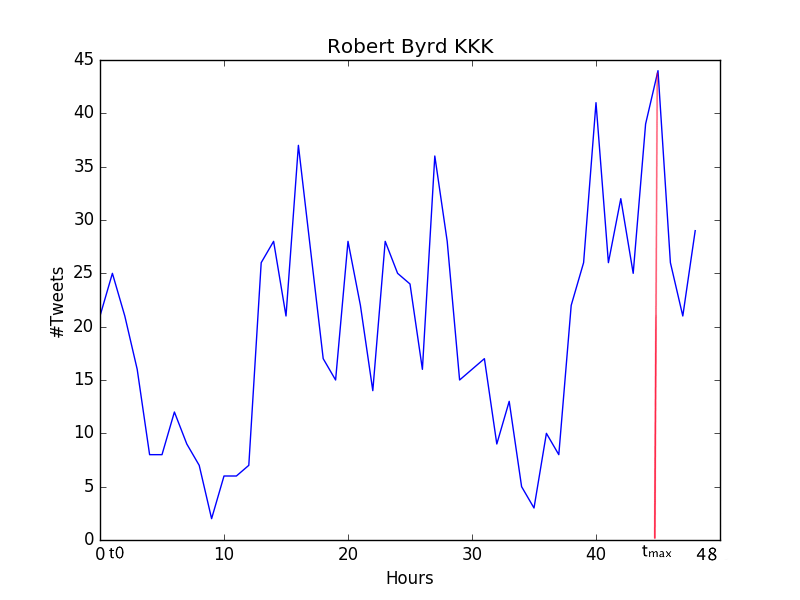}
}
\caption{tweet volume of the rumor event of Robert Byrd at full scale and after selected time period}
\label{fig:KKK_part}
\end{figure}

\subsection{Single Tweet Classification Experiments} 
For this task, we developed two kinds of classification models: traditional classifier with handcrafted features and neural networks without tweet embeddings. For the former, we used 27 distinct surface-level features extracted from single tweets (analogously to the Twitter-based features presented in Section~\ref{sub:features}). For the latter, we select the baselines from state-of-the-art text classification models, i.e., Basic tanh-RNN~\cite{madetecting}, 1-layer GRU-RNN~\cite{madetecting}, 1-layer LSTM~\cite{madetecting}, 2-layer GRU-RNN~\cite{madetecting}, FastText~\cite{joulin2016bag} and CNN+LSTM~\cite{zhou2015c} model. The hybrid model CNN+LSTM is adapted in our work for tweet classification.  

\begin{figure}[!h]

\centering

\subfigure[Sample RNN Model]{\label{fig:SRNN}
\centering
  \includegraphics[width=0.12\columnwidth]{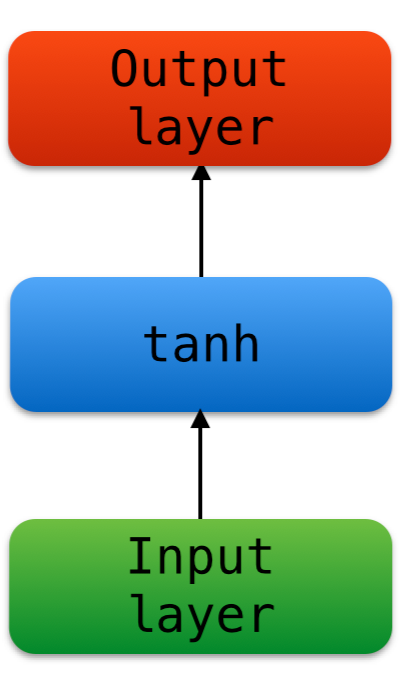}
} %
\subfigure[1-layer GRU-RNN]{\label{fig:1GRU}
\centering
  \includegraphics[width=0.12\columnwidth]{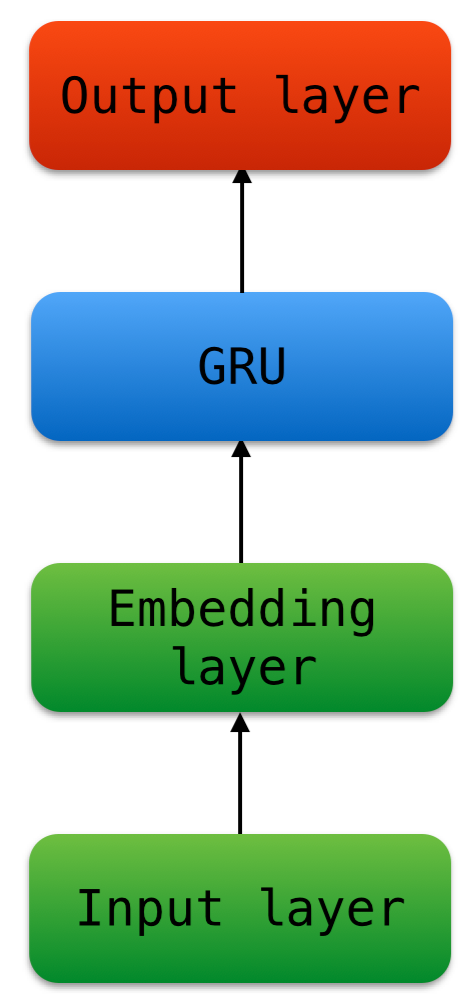}
}
\subfigure[1-layer LSTM-RNN]{\label{fig:1LSTM}
\centering
  \includegraphics[width=0.12\columnwidth]{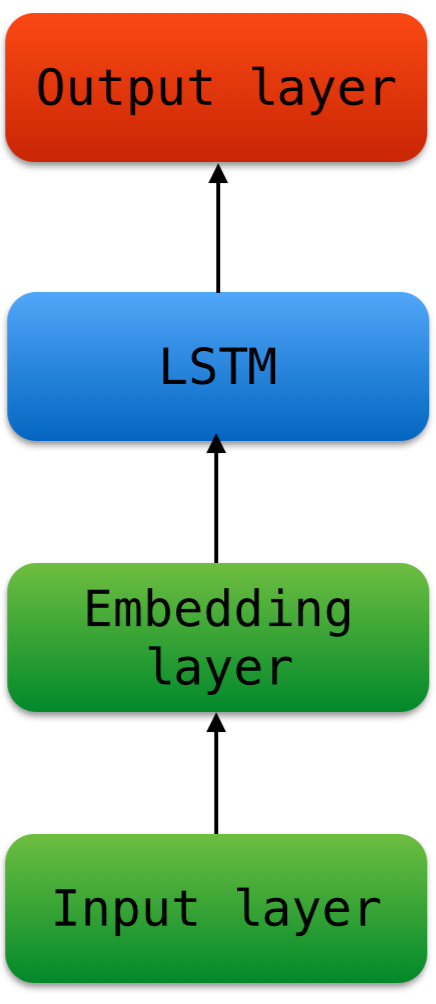}
}
\subfigure[2-layer GRU-RNN]{\label{fig:2GRU}
\centering
  \includegraphics[width=0.12\columnwidth]{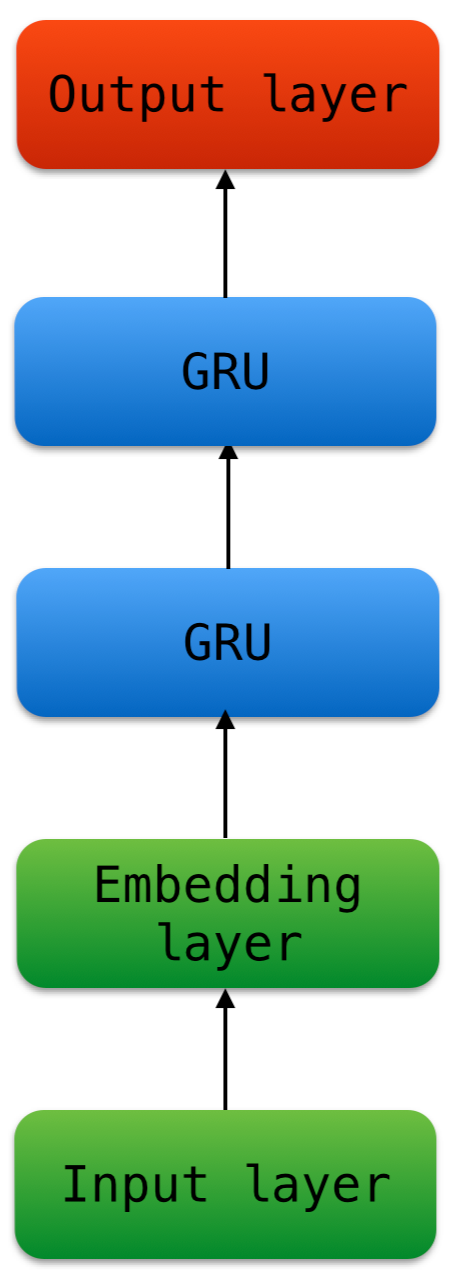}
}
\subfigure[FastText]{\label{fig:fasttext}
\centering
  \includegraphics[width=0.12\columnwidth]{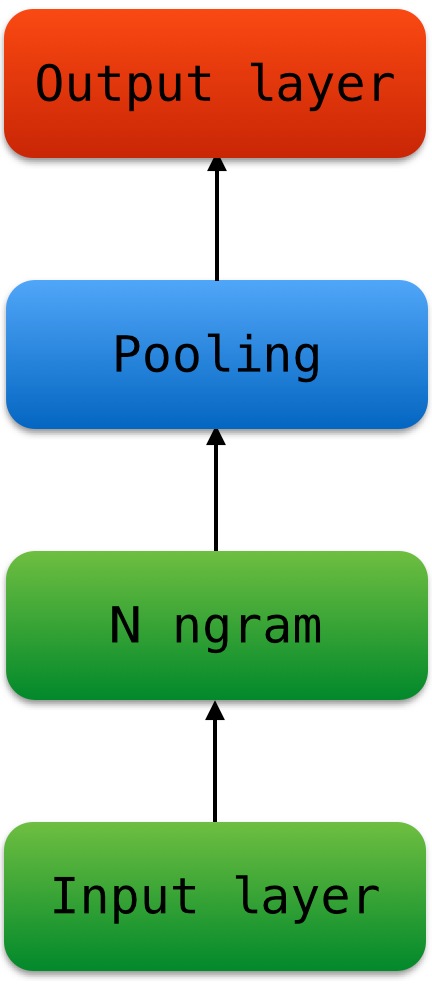}
}
\subfigure[Hybrid CNN+LSTM]{\label{fig:CNNLSTM}
\centering
  \includegraphics[width=0.12\columnwidth]{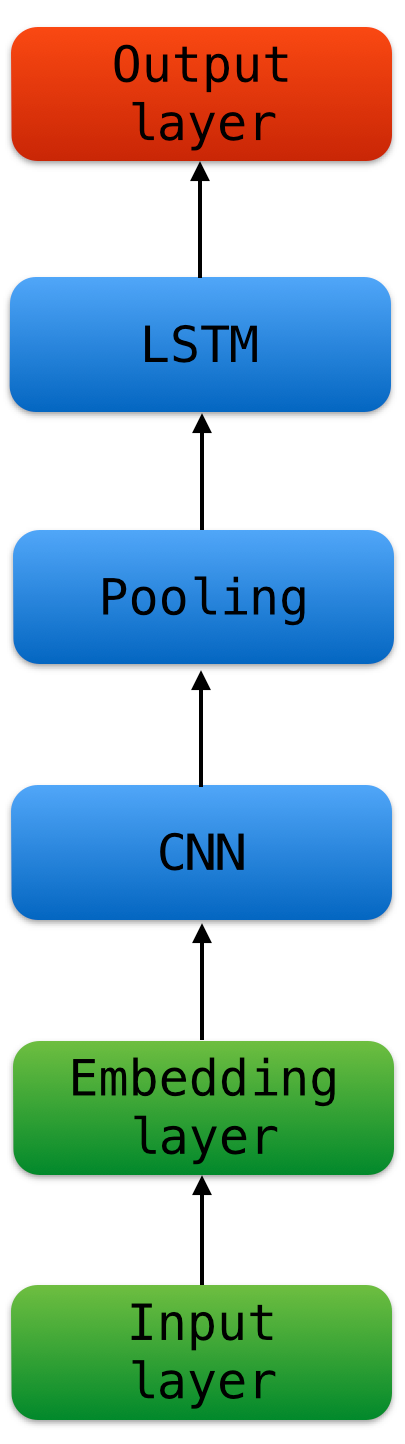}
}
\caption{RNN-based Models for Single Tweet Classification}
\label{fig:NNModel1}
\end{figure}

\subsubsection{Single Tweet Model Settings.} For the evaluation, we shuffle the 180 selected events and split them into 10 subsets which are used for 10-fold cross-validation (we make sure to include near-balanced folds in our shuffle). For the experiments, we implement the 3 non-neural network models with Scikit-learn library\footnote{scikit-learn.org/}. Furthermore, we implement the neural network with TensorFlow~\footnote{https://www.tensorflow.org/} and Keras\footnote{https://keras.io/}. The first hidden layer is an embedding layer, which is set up for all tested models with the embedding size of 50. The output of the embedding layer are low-dimentional vectors representing the words. To avoid overfitting, we use the 10-fold cross validation and dropout for regularization with dropout rate of 0.25.

%
%
\subsection{Single Tweet Classification Results}  
The experiments' results of the testing models are shown in Table \ref{tab:single_result}. The best performance is achieved by the CNN+LSTM model with a good accuracy of 81.19\%. The non-neural network model with the highest accuracy is RF. However, it reaches only 64.87\% accuracy and the other two non-neural models are even worse. So the classifiers with hand-crafted features are less adequate to accurately distinguish between rumors and news. For analysing the employed handcrafted features, we rank them by importances using RF (see \ref{tab:Features_Importance}). It can be seen that the best feature is the sentimental polarity scores, a high-level text-based feature.
 

  \begin{minipage}{\columnwidth}
  \begin{minipage}[b]{0.45\columnwidth}
      \centering
    \begin{adjustbox}{center, width=0.8\columnwidth} 
   \begin{tabular}{@{}lllllll@{}}
 \toprule
 \textbf{Model} & \textbf{Accuracy} \\ \midrule
  \textbf{CNN+LSTM} & \textbf{0.8119 }\\
 2-layer GRU & 0.7891\\
 GRU & 0.7644\\
 LSTM & 0.7493\\
 Basic RNN with tanh &  0.7291\\
 FastText &  0.6602\\ \bottomrule
 Random Forest & \textbf{0.6487 }\\
 SVM &  0.5802\\
 Decision Trees &  0.5774\\ \bottomrule
 \end{tabular}
 \end{adjustbox}
\captionof{table}{Performance of different credibility models}
 \label{tab:single_result}

 \end{minipage}
  \hfill
  \begin{minipage}[b]{0.45\columnwidth}
    \centering
    \begin{adjustbox}{center, width=0.7\columnwidth}   
 \begin{tabular}{@{}lllllll@{}}
\toprule
\textbf{Feature} & \textbf{Importance} \\ \midrule
PolarityScores	&	0.146\\
Capital	&	0.096\\
LengthOfTweet  &	0.092\\
UserTweets  &	0.087 \\
UserFriends  &	0.080 \\
UserReputationScore  &	0.080 \\
UserFollowers   &	0.079 \\
\bottomrule

\end{tabular}
\end{adjustbox}
 \captionof{table}{Top Features Importance}
\label{tab:Features_Importance}

\end{minipage}
 \end{minipage}

\textbf{Discussion of Feature Importance}  
For analysing the employed features, we rank them by importances using RF (see \ref{tab:Features_Importance}). The best feature is related to sentiment polarity scores. There is a big bias between the sentiment associated to rumors and the sentiment associated to real events in relevant tweets. In specific, the average polarity score of news event is -0.066 and  the average of rumors is -0.1393, showing that rumor-related messages tend to contain more negative sentiments. Furthermore, we would expect that verified users are less involved in the rumor spreading. However, the feature appears near-bottom in the ranked list, indicating that it is not as reliable as expected. Also interestingly, the feature``IsRetweet'' is also not as good a feature as expected, which means the probability of people retweeting rumors or true news are similar (both appear near-bottom in the ranked feature list).It has to be noted here that even though we obtain reasonable results on the classification task in general, the prediction performance varies considerably along the time dimension. This is understandable, since tweets become more distinguishable, only when the user gains more knowledge about the event.

\subsection{Rumor Datasets and Model Settings} 
  We use the same dataset described in Section \ref{sec:dataset_single}. In total --after cutting off 180 events for pre-training single tweet model -- our dataset contains 360 events and 180 of them are labeled as rumors. As a rumor is often of a long circurlating story~\cite{friggeri2014rumor}, this results in a rather long time span. In this work, we develop an event identification strategy that focuses on the first 48 hours after the rumor is peaked. We also extract 11,038 domains, which are contained in tweets in this 48 hours time range.
  

\begin{table}[!h]
 \centering
\scalebox{0.8}{
 \begin{tabular}{@{}cccccc@{}}
 \toprule
 \textbf{Type} & \textbf{Earliest Event} & \textbf{Latest Event}&  \textbf{Average Time Span (days)} \\ \midrule
 News & 21.02.2012 & 16.07.2016  &  4.5\\ \midrule
 Rumors & 20.10.2009  & 04.08.2016& 1679.6 \\
 	 \bottomrule
 \end{tabular}}
 \caption{Time Span of News and Rumors}
 \label{tab:Tweet_Time}
\end{table}

\textbf{Settings.} For the time series classification model, we only report the best performing classifiers, SVM and Random Forest, here. The parameters of SVM with RBF kernel are tuned via grid search to $C=3.0$, $\gamma = 0.2$. For Random Forest, the number of trees is set to be 350. All models are trained using 10-fold cross validation. 


 \subsection{Rumor Classification Results} 
 Here we present the experiment results. We test all models by using 10-fold cross validation, the experiment result is shown in the Table \ref{tab:time_result}. TS-RF is the time series model learned with Random Forest of all features, $TS-SVM$ is the baseline from~\cite{ma2015detect}, $TS-SVM_{all}$ is $TS-SVM$ with all features, $TS-SVM_{Credit}$ is $TS-SVM$ with CreditScore, $TS-SVM_{SpikeM}$ is $TS-SVM$ with epidemiological features. In the lower part of the table, $RNN_{el}$ is the RNN model at event-level~\cite{madetecting}. As shown in the Table \ref{tab:time_result} and as targeted by our early detection approach, our model has the best performance in all cases over the first 24 hours, remarkably outperforming the baselines in the first 12 hours of spreading. The performance of $RNN_{el}$ is relatively low, as it is based on aggregated \textit{contents}. This is expected as the news (non-rumor) dataset used in ~\cite{madetecting} are crawled also from SNOPES.com, in which events are often of small granularity (aka. subless). As expected, exploiting contents solely at event-level is problematic for high-impact, evolving events on social media. As the main aim of the work is to give a wide understanding of the feature contributions over the early-stage time period, we did not experiment our approach in the same dataset with~\cite{madetecting}; and leave a deeper investigation on the sub-event issue to future work. As shown in the Table \ref{tab:time_result}, it is interesting to see that, the model \textbf{TS-RF} that engages a wide range of features, has best performance in all cases over time and \textsf{CreditScore} can significantly improve the performance of time series model in the first 24 hours of event. And it is at least the second best feature in all cases over time.

\begin{table}[]
\centering
\scalebox{0.7}{
\begin{tabular}{|c|ccccccccc|}
\hline
\multicolumn{1}{|c|}{\multirow{2}{*}{Model}} & \multicolumn{9}{c|}{Accuracy in hours}                     \\ \cline{2-10} 
\multicolumn{1}{|l|}{}& 1 & 6 &12& 18 & 24 & 30 & 36  & 42 & \multicolumn{1}{c|}{48} \\\hline
 $TS-RF$  & \textbf{0.82} &  \textbf{0.84} &  \textbf{0.84} &\underline{0.84} & \textbf{0.87}& \underline{0.87} &\textbf{0.88}&\underline{0.89}&\textbf{0.91 }\\
$TS-SVM_{all}$ &\underline{0.76} & 0.79  &  \underline{0.83} & 0.83 &   \underline{0.87} &   \textbf{0.88}&   0.86&   0.89 & \underline{0.90}  \\

$TS-SVM_{Credit}$& 0.73 & \underline{0.80}  & 0.83  & \textbf{0.85} &   0.85 & 0.86  &\underline{0.88} &\textbf{0.90} &\underline{0.90}\\

$TS-SVM_{Epi}$&0.70 & 0.77 & 0.81  & 0.81 &  0.84 &0.83  &0.81 &0.85 & 0.86\\

$TS-SVM_{SpikeM}$&0.69 & 0.74 & 0.81  & 0.80 &  0.85 &0.86 &0.86 &0.86  & 0.86\\
$TS-SVM$~\cite{ma2015detect} &0.69  &0.76 & 0.81& 0.81 &  0.84   &0.86 & 0.87& 0.88 & 0.88  \\
\hline
$RNN_{el}$~\cite{madetecting}&0.68 & 0.77 & 0.81  & 0.81 &  0.84 &0.83  &0.81 &0.85 & 0.86\\
$SVM_{static}+Epi$~\cite{jin2013epidemiological} &0.60& 0.69  & 0.71&0.72   &  0.75 &0.78& 0.75&0.78&0.81 \\   

$SVM_{static}+SpikeM$~\cite{kwon2013prominent}&0.58& 0.68  & 0.72&0.73   &  0.77 &0.78& 0.78&0.79&0.77 \\   

$SVM_{static}$~\cite{yang2012automatic} &0.62& 0.70  & 0.70&0.72   &  0.75 & 0.80& 0.79&0.78&0.77 \\   

\bottomrule           
\end{tabular}
}
 \caption{Performance of different models over time, numbers in Bold is best accuracy, in underlined is second-to-best. \textsf{TS} indicates time-series structure.}
 \label{tab:time_result}
\end{table}
%
%
%


 \subsubsection{Feature Analyzing Over Time} 
 \label{featureanalyzing}
 Here we present the performance of features over time. We use the RF permutation-based (that account for possible feature correlations) for measuring feature importance. First we split the features in 7 catalogues as in Table \ref{tab:full_features}: \emph{Tweet\_Feature}, \emph{User\_Feature}, \emph{Text\_Feature},  \emph{CreditScore}, \emph{SpikeM Features}, \emph{Epidemiological Features}, \emph{CrowdWisdom} and the \emph{BestSet}. The \emph{BestSet} is a combination of the top 9 most important features which is mentioned in the below paragraph. The results over 48 hours are in Figure \ref{fig:allfeature} .

 \begin{figure}[!h]
\centering
\includegraphics[width=\columnwidth]{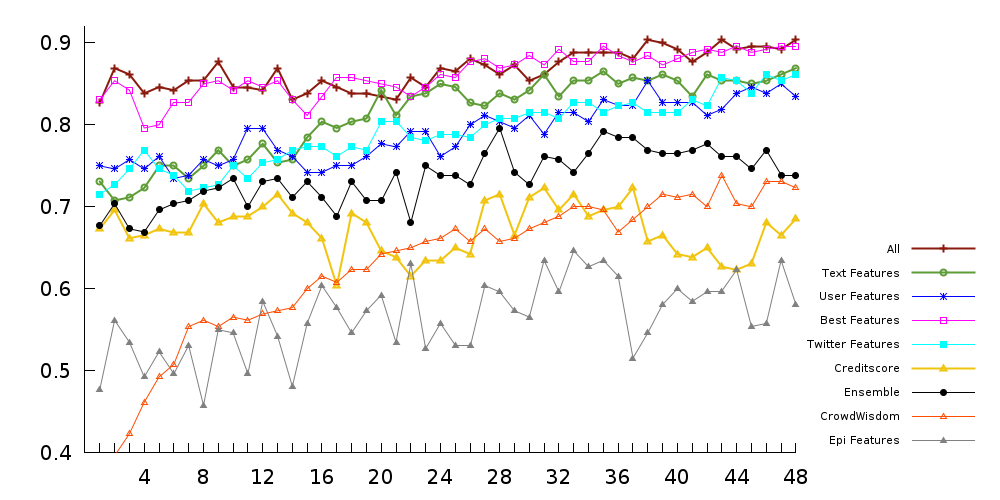}
\caption{Accuracy: Performance of different feature groups over time}
\label{fig:allfeature}
\end{figure}
 As we can see in Figure \ref{fig:allfeature} the best result on average over 48 hours is the \emph{BestSet}. Second one is \emph{All features}. Except those two, the best group feature is \emph{Text features}. One reason is the text feature set has the largest group of feature with totally 16 features. But if look into each feature in text feature group, we can see the best and the worst features are all in this set. \emph{User features} and \emph{Twitter features} are stable over time around 82\%. The performances of 3 different models (SIS, SEIZ and SpikeM) describing the propagation pattern of rumors and news are not ideal especially within 24 hours. \emph{CrowdWisdom} and \emph{CreditScore} both contain only one feature, but they already have impressive results comparing with the \emph{User features} and \emph{Twitter features}.
  

%
%
%
%
%
%
%
 
 \subsubsection{Text Features} 
 \emph{Text feature set} contains totally 16 features. The feature ranking are shown in Table \ref{testfeaturerank}. The best one is \emph{NumOfChar} which is the average number of different characters in tweets. \emph{PolarityScores} is the best feature when we tested the single tweets model, but its performance in time series model is not ideal. It is true that rumor contains more negative sentiment, but in an event (rumor or news) people can show their mixed views about this event \cite{mendoza2010twitter,starbird2014rumors} like discussing or denying, so the \emph{PolarityScores}'s performance becomes worse over time. \emph{Text features} overall are shown to be more effective than \textit{Twitter} and \textit{user} feature sets.
 
 \begin{table}[!h]
 \centering
\scalebox{0.7}{
\begin{tabular}{@{\textbf{ }}ccccccccccccccccc@{}}
\toprule
\textbf{Features} & \multicolumn{10}{c}{\textbf{Ranks}} \\\hline
Hours & 1 & 6 & 12 & 18&24&30&36&42&48 & AVG\\\hline
NumOfChar			& 3& 3 & 4 & 3& 3& 8& 7&6 & 4& 4.29\\
QuestionExclamation		& 25& 16 & 2 & 1& 1& 1& 3& 7&5&4.79\\
PolarityScores & 12 & 15& 23 & 28 & 33& 33& 34& 31&32&28\\

Smile 			& 35 & 45& 45 & 48 & 48& 48& 48& 48&48&47.06\\ 
 
Sad 			& 36 & 46& 46 & 49 & 49& 49& 49& 49&49&47.9\\ 

\bottomrule
 \end{tabular}}
\caption{Importance ranking of text features over time.}
\label{testfeaturerank}
\end{table}
 \subsubsection{Twitter Features} 
   The performance of \emph{Twitter features} are stable over time from the beginning to the end. The 3 best of \emph{Twitter Features} are all based on contained URLs in tweets: \emph{ContainNEWS}, \emph{UrlRankIn5000}, \emph{WotScore}, as shown in Table \ref{twitterfeaturerank}. It is quite reasonable that the news event would have higher probability to be reported by news or authorized websites. And it is clear to see that their performances significantly improve after 24 hours. But the other original Twitter functions like the retweets or mention do not contribute much.

\begin{table}[!h]
\centering
\scalebox{0.7}{
\begin{tabular}{@{\textbf{ }}ccccccccccccccccc@{}}
\toprule
\textbf{Features} & \multicolumn{10}{c}{\textbf{Ranks}} \\\hline
Hours & 1 & 6 & 12 & 18&24&30&36&42&48 & AVG\\\hline
ContainNEWS & 8 & 4 & 5& 4 & 4&2&2&2&2&3.48\\
WotScore			& 4 & 10& 6 & 10 & 10& 6& 9& 8&7&7.63\\ 
Mention 			& 13 & 5& 10 & 14 & 13& 10& 12& 12&10&10.98\\
Hashtag 			& 20 & 20& 15 & 18 & 16& 13& 15& 17&17&17.46\\ 
Retweets & 21 & 21& 27 & 38 & 42& 35& 31& 37&34&33.25\\
\bottomrule
 \end{tabular}}
\caption{Importance ranking of Twitter features over time.}
\label{twitterfeaturerank}
\end{table}

\subsubsection{User Features} 
The performance of \emph{user features} is similar with the \emph{Twitter features}, they are both quite stable from the first hour to the last hour. As shown in Table \ref{userfeaturerank}, the best feature over 48 hours of the user feature group is \emph{UserTweetsPerDays} and it is the best feature overall in the first 4 hours, but its rank decreases with time going by. Others user-based features like \textit{UserReputationScore} and \textit{UserJoinDate} also have a better performance in the first fews hours. That means the sources (the posters in the first few hours) of news and rumors are quite different with each other. But with more and more users joining in the discussion, the bias of two groups of users becomes less. After 6 hours, it seems that we can better distinguish the rumors based on the tweet contents \emph{(text features)}, rather than relying on the features of users.
 
\begin{table}[!h]
\centering
\scalebox{0.7}{
\begin{tabular}{@{\textbf{ }}ccccccccccccccccc@{}}
\toprule
\textbf{Features} & \multicolumn{10}{c}{\textbf{Ranks}} \\\hline
Hours & 1 & 6 & 12 & 18&24&30&36&42&48 & AVG\\\hline
UserTweetsPerDays & 0 & 1 & 1& 2 & 2&9&5&10&14&4.63\\
UserReputationScore	 & 1& 2 & 3 & 5 &6& 5& 6& 3&3&5.06\\
UserJoin\_date			& 5 & 8& 8 & 8 & 12& 14& 16& 11&9&10.58\\ 
UserVerified 			& 24 & 17& 12 & 16 & 17& 12& 11&14&19&16.25\\ 
 \bottomrule
 \end{tabular}}
\caption{Importance ranking of user features over time.}
\label{userfeaturerank}
\end{table}
\subsubsection{Epidemiological Features}
The performance of this feature group is not so convincing. The feature $P_a$ from SpikeM model is the best one of them. The problem of these two models which we have already figured out in Section \ref{sec:epide} is that two models need substantial data to fit the parameters. After 24 hours, model trained with these epidemiological featuresreaches 60\% in accuracy. In other words, before 24 hour these is no clear propagation pattern of these events. In~\cite{kwon2013prominent}, the durations of dataset are more than 60 days. In~\cite{jin2013epidemiological}, they use 160 hours' tweets' volume to fit the SEIZ models. Their event durations are far larger than ours focused 48 hours. The $P_a$ parameter from SpikeM is the only feature barely has some contributions for rumor detection in our experiment. It stands for the strength of periodicity in SpikeM.~\cite{kwon2013prominent} add 3 more parameters $Q_a$,$Q_p$ and $Q_s$ to explain the periodicity of the external shock, but they do not produce same effect in our experiment, because 48 hours time period is rather too short to contain multi-pike patterns.

\begin{table}[!h]
\centering
\scalebox{0.7}{
\begin{tabular}{@{\textbf{ }}ccccccccccccccccc@{}}
\toprule
\textbf{Features} & \multicolumn{10}{c}{\textbf{Ranks}} \\\hline
Hours & 1 & 6 & 12 & 18&24&30&36&42&48 & AVG\\\hline
$P_a$ & 29 & 28 & 34& 30 & 33&24&23&21&23&25.75\\
$R_{SI}$	 & 47& 24 & 30 & 23 &36& 39& 38& 24&30&29.56\\
$\beta_{SIS}$			&49& 30 & 33& 28 & 31 & 36&28&33&25&30.15\\ 
$Q_a$  			& 44 & 47& 47 & 21 & 38& 40& 44&41&33&38.04\\ 
 \bottomrule
 \end{tabular}}
 \caption{Importance ranking of Epidemiological Features}
\label{SPikemfeaturerank}

\end{table}
 
 \subsubsection{CreditScore and CrowdWisdom}. As shown in Table~\ref{tab:Rank_Credit}, \emph{CreditScore} is the best feature in general. Figure \ref{fig:WCVSAF} shows the result of models learned with the full feature set with and without \emph{CreditScore}. Overall, adding \textit{CreditScore} improves the performance, significantly for the first 8-10 hours. The performance of \textit{all-but-CreditScore} jiggles a bit after 16-20 hours, but it is not significant. \emph{CrowdWisdom} is also a good feature which can get 75.8\% accuracy as a single feature. But its performance is poor (less than 70\%) in the first 32 hours getting better over time (see Table~\ref{tab:Rank_Credit}). Table~\ref{tab:Rank_Credit} also shows the performance of \textit{sentiment} feature (\textit{PolarityScores}), which is generally low. This demonstrates the effectiveness of our \textit{curated} approach over the \textit{sentiments}, yet the crowd needs time to unify their views toward the event while absorbing different kinds of information. 

 \begin{figure}[!h]
\centering
\includegraphics[width=0.8\columnwidth]{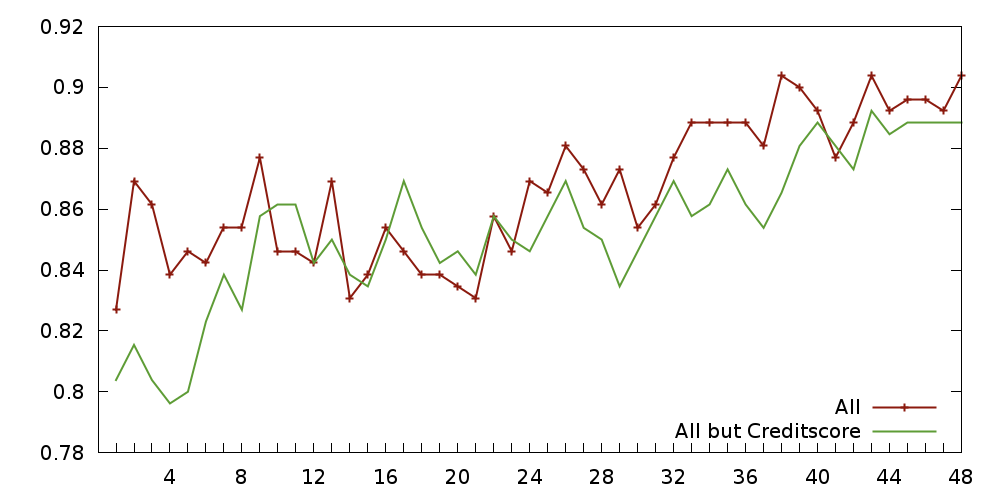}
\caption{Accuracy: All features vs without CreditScore}
\label{fig:WCVSAF}
\end{figure}

%
  
  \begin{table}[!h]
\centering
\scalebox{0.7}{
\begin{tabular}{@{\textbf{ }}ccccccccccccccccc@{}}
\toprule
\textbf{Features} & \multicolumn{10}{c}{\textbf{Ranks}} \\\hline
Hours & 1 & 6 & 12 & 18&24&30&36&42&48 & AVG\\\hline
CreditScore & 1 & 0 & 0& 0 & 0&0&0&0&0&0.08\\
CrowdWisdom	 & 34& 38 & 21 & 14& 8& 5& 5& 2&2&13.18\\
 \bottomrule
 \end{tabular}}
 \caption{Ranks of CreditScore and CrowdWisdom}
\label{tab:Rank_Credit}

 \end{table}

\subsubsection{Machine vs Human} 
In this section, we compare the performance our model with the human rumor debunking websites: \textbf{snopes.com} and  \textbf{urbanlegend.com}. \textit{Snopes} has their own Twitter account\footnote{https://twitter.com/snopes}. They regularly post tweets via this account about rumors which they collected and verified. We consider the creation time of the first tweet which mentions \emph{"@snopes"} or contains \emph{"urbanlegend"} in the text or in URLs is the timestamp of human debunking the rumors. However, a rumor-related news tends to have a long discussion over time, making this determination is hard. For example, the rumor about the rapper Tupac Shakur, who is thought to have been killed in 1996, is still alive and comes out of hiding. This topic bursted in 2012, 2015 and 2016 several times and the tweets' volume of 2012 is the highest peak. But Snopes reported this rumor in the september 2015\footnote{http://www.snopes.com/media/notnews/tupac.asp}. So we consider that they don't refer to the same rumor affair.
  \begin{figure}[!h]
\centering
\includegraphics[width=0.8\columnwidth]{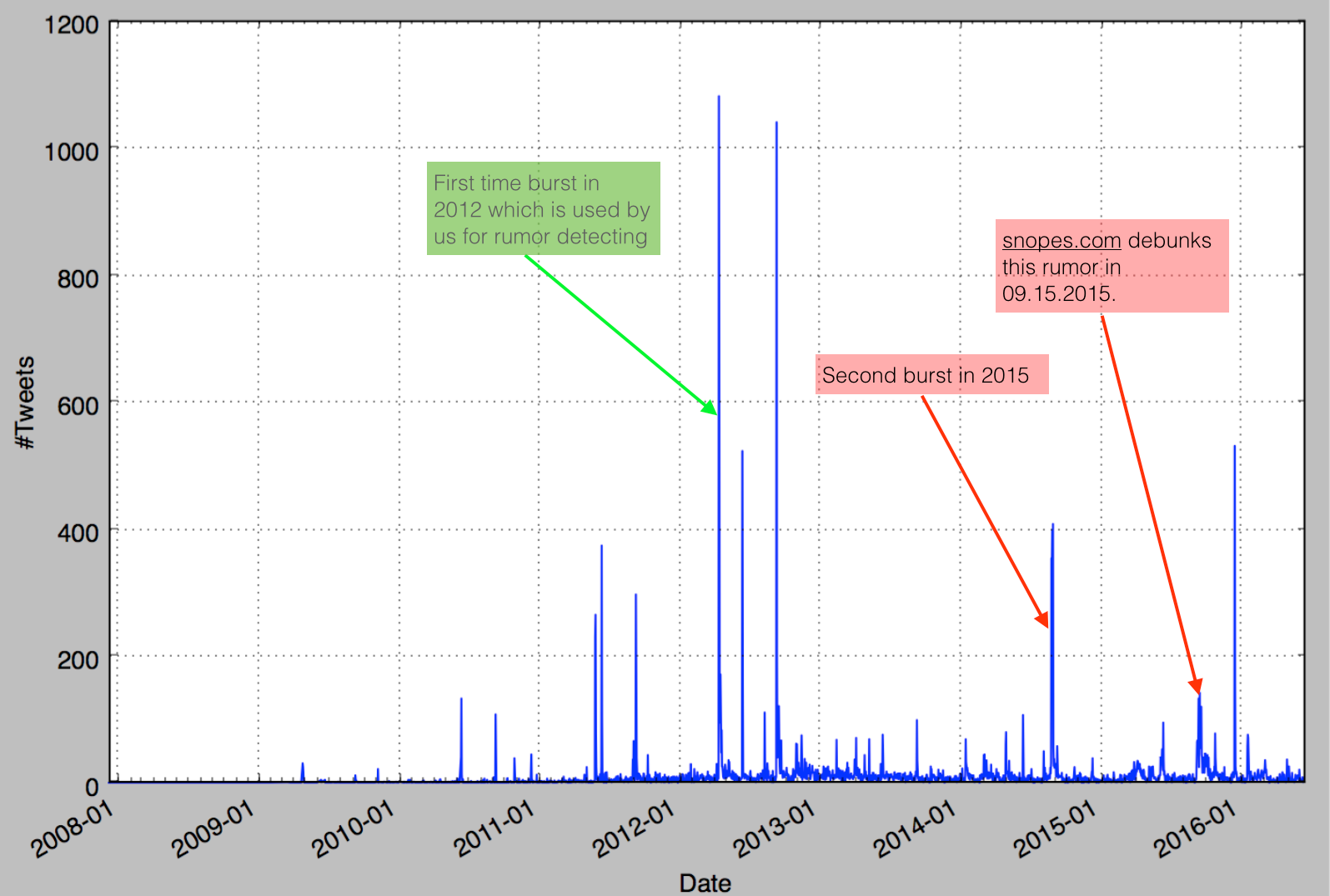}
\caption{Tweet Volume Of Rumor about Tupac Shakur}
\label{fig:Multipike}
\end{figure}   
   
To overcome this issue, we set up a threshold 72 hours. We only consider the first candidate within 72 hours before or after the beginning time of the event as timestamp of human confirming rumors. On average the human editors of Snopes need 25.49 hours to verify the rumors and post it. Our system already achieves 87\% accuracy in 25 hours. We illustrate two examples here in Figures~\ref{fig:ealiset_rumo} and~\ref{fig:lastest_rumo}. Figure \ref{fig:ealiset_rumo} is a rumor about `Okra curing diabetes' \footnote{http://www.snopes.com/medical/homecure/okra.asp} which we detected the beginning time is 01.31.2014 04:00. Snope debunked it at 01.28.2014 21:00, 55 hours earlier than our study time period. However, Snopes does not provide any information regarding how they detect the rumor. Figure \ref{fig:lastest_rumo} depicts another example, showing that human detect it 71 hour after the event starts, which is the latest detection in our study. Despite those issues, we show the comparision results in Table \ref{tab:Human_confit_rumor}.
 
     \begin{figure}[!h]
\centering
\subfigure[Earliest]{\label{fig:ealiset_rumo}
\includegraphics[width=0.4\columnwidth]{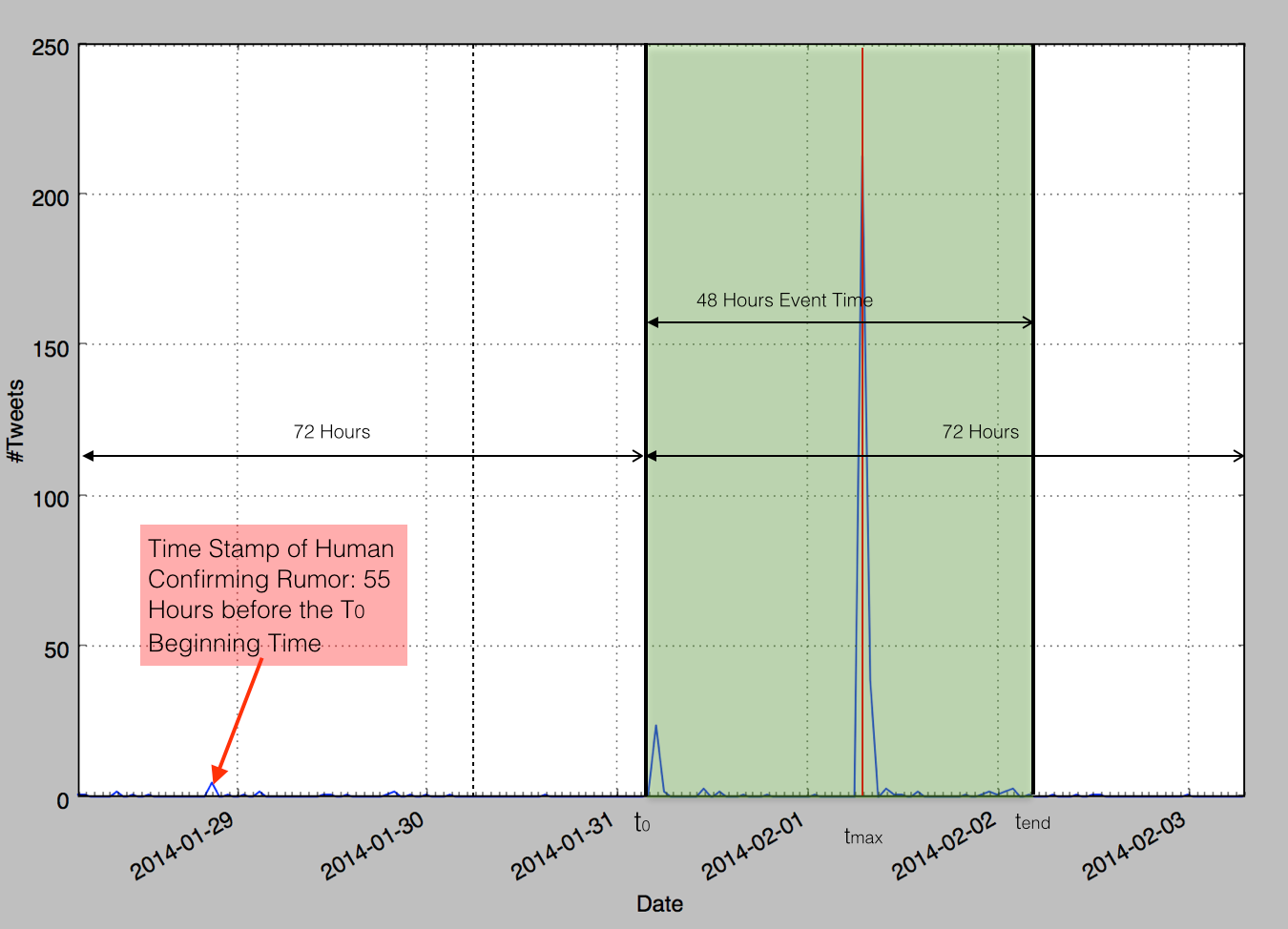}
}
\subfigure[Latest]{\label{fig:lastest_rumo}
\includegraphics[width=0.4\columnwidth]{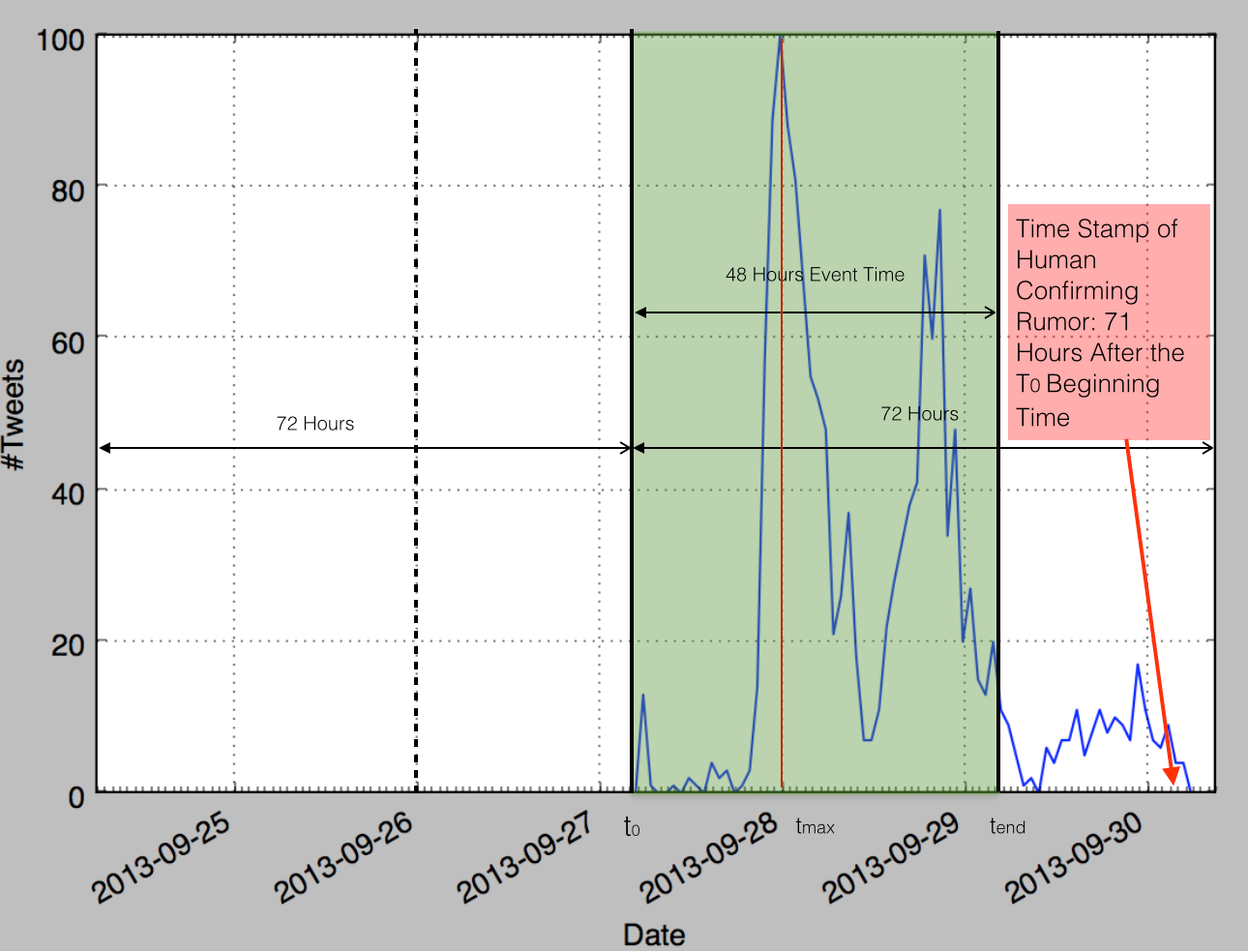}
}
\caption{The Earliest and Latest Time Stamp Of Human Confirming Rumor}
\end{figure} 

  \begin{table}[!h]
 \centering
\scalebox{0.7}{
\begin{tabular}{@{\textbf{ }}cll@{}}
\toprule
\textbf{} & \textbf{Hours} \\ \midrule
Latest Time of Human Detection & 71\\
Earliest Time of Human Detection			& -55\\
Average 		& 25.49\\
\bottomrule

\end{tabular}}
\caption{Time of Human Confirming Rumors }
\label{tab:Human_confit_rumor}
\end{table}

\subsubsection{Case Study: Munich Shooting}. 
We first show the event timeline at an early stage. Next we discuss some examples of  misclassifications by our ``weak'' classifier and show some analysis on the strength of some highlighted features. The rough event timeline looks as follows.
\begin{small}
\begin{itemize}
\item At 17:52 CEST, a shooter opened fire in the vicinity of the Olympia shopping mall in Munich. 10 people, including the shooter, were killed and 36 others were injured. 

\item At 18:22 CEST, the first tweet was posted. There might be some certain delay, as we retrieve only tweets in English and the very first tweets were probably in German. The tweet is \emph{"Sadly, i think there's something terrible happening in \#Munich \#Munchen. Another Active Shooter in a mall. \#SMH"}. 

\item At 18:25 CEST, the second tweet was posted: \emph{"Terrorist attack in Munich????"}.   

\item At 18:27 CEST, traditional media (BBC) posted their first tweet. \emph{"'Shots fired' in Munich shopping centre - http://www.bbc.co.uk/news/world-europe-36870800a02026 @TraceyRemix gun crime in Germany just doubled"}.

\item At 18:31 CEST, the first misclassified tweet is posted. It was a tweet with shock sentiment and swear words: \emph{"there's now a shooter in a Munich shopping centre.. What the f*** is going on in the world. Gone mad"}. It is classified as \textit{rumor-related}.
\end{itemize}
\end{small}
We observe that at certain points in time, the volume of rumor-related tweets (for sub-events) in the event stream surges. This can lead to \textit{false positives} for techniques that model events as the aggregation of all tweet contents; that is undesired at critical moments. We trade-off this by debunking at single tweet level and let each tweet vote for the credibility of its event. We show the \emph{CreditScore} measured over time in Figure \ref{fig:munichattackCS}. It can be seen that although the credibility of some tweets are low (rumor-related), averaging still makes the \emph{CreditScore} of \textsf{Munich shooting} higher than the average of news events (hence, close to a \textit{news}). In addition, we show the feature analysis for ContainNews (percentage of URLs containing news websites) for the event \textit{Munich shooting} in Figure \ref{fig:munichattackNews}. We can see the  curve of \textit{Munich shooting} event is also close to the curve of average news, indicating the event is more news-related.

\begin{figure}[h!]
\subfigure[CreditScore 12 hours]{\label{fig:munichattackCS}
\centering
\includegraphics[width=0.45\columnwidth]{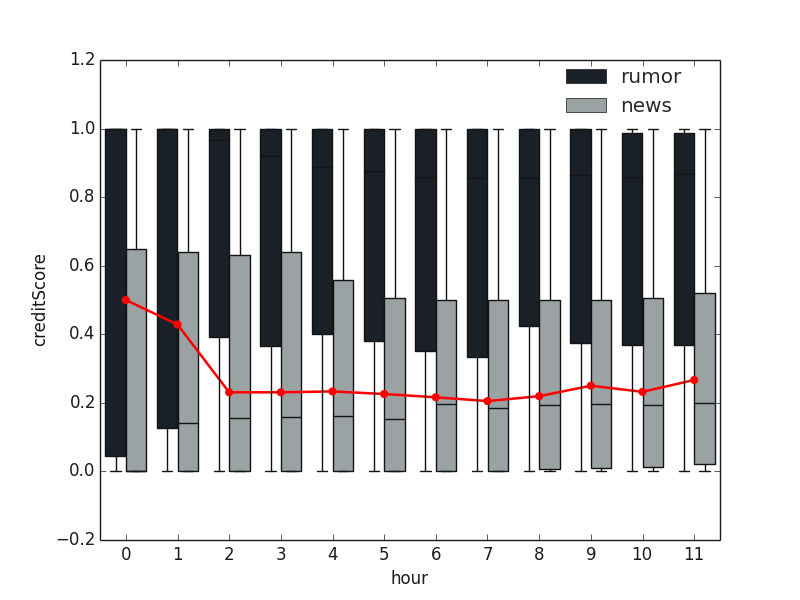}
}
\subfigure[ContainsNews 12 hours]{\label{fig:munichattackNews}
\centering
\includegraphics[width=0.45\columnwidth]{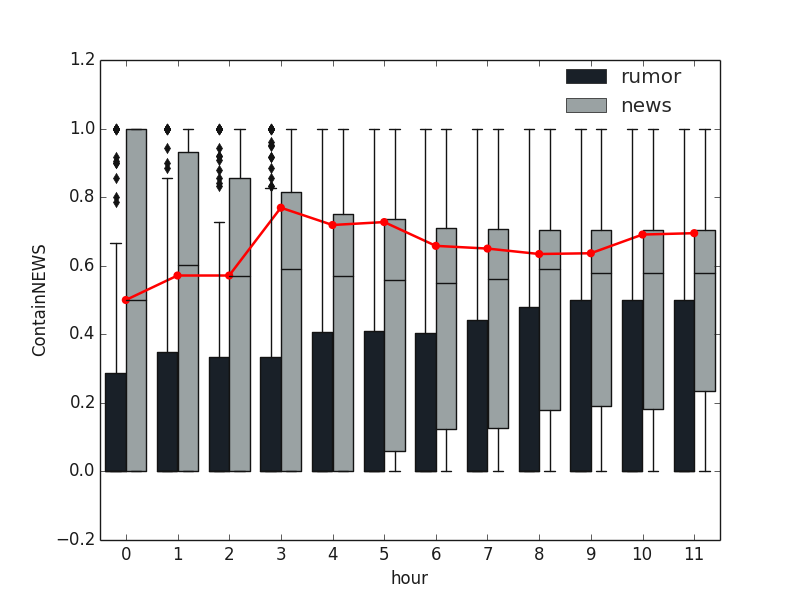}
}
\subfigure[CreditScore 12 hours]{\label{fig:munichattackCS2}
\centering
\includegraphics[width=0.45\columnwidth]{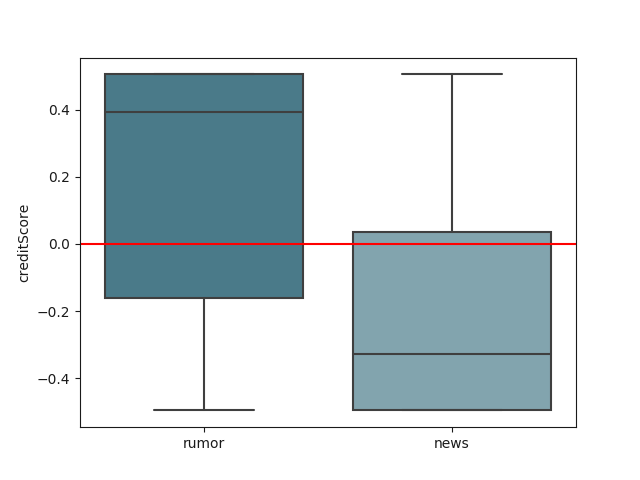}
}
\subfigure[ContainsNews 12 hours]{\label{fig:munichattackNews2}
\centering
\includegraphics[width=0.45\columnwidth]{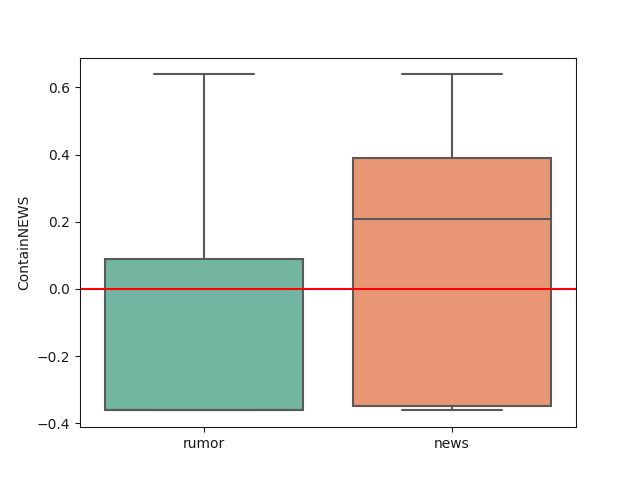}
}
\subfigure[CreditScore 48 hours]{\label{fig:munichattackCS2}
\centering
\includegraphics[width=0.45\columnwidth]{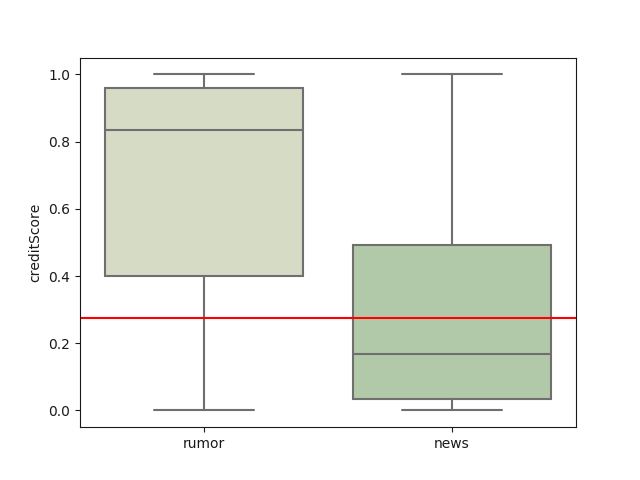}
}
\subfigure[ContainsNews 48 hours]{\label{fig:munichattackNews2}
\centering
\includegraphics[width=0.45\columnwidth]{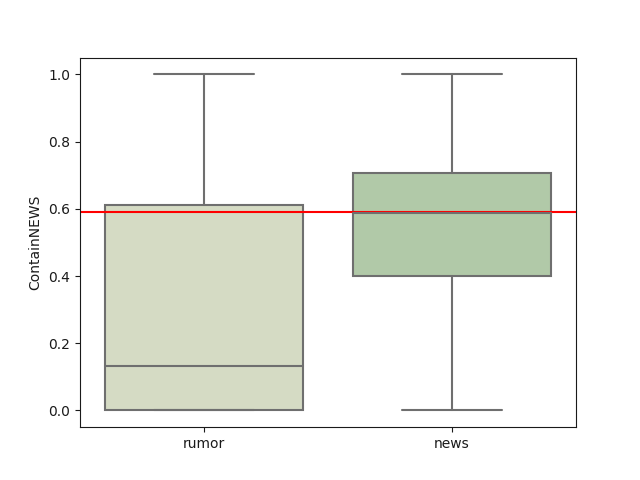}
}
\subfigure[Polarity 48 hours]{\label{fig:munichattackCS2}
\centering
\includegraphics[width=0.45\columnwidth]{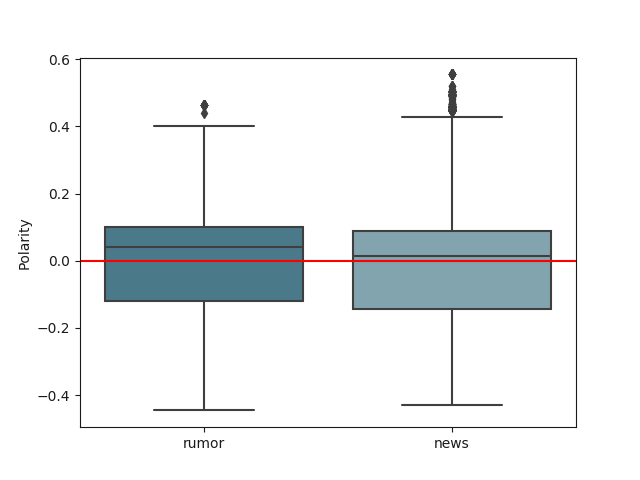}
}
\subfigure[Users in large cities 48 hours]{\label{fig:munichattackNews2}
\centering
\includegraphics[width=0.45\columnwidth]{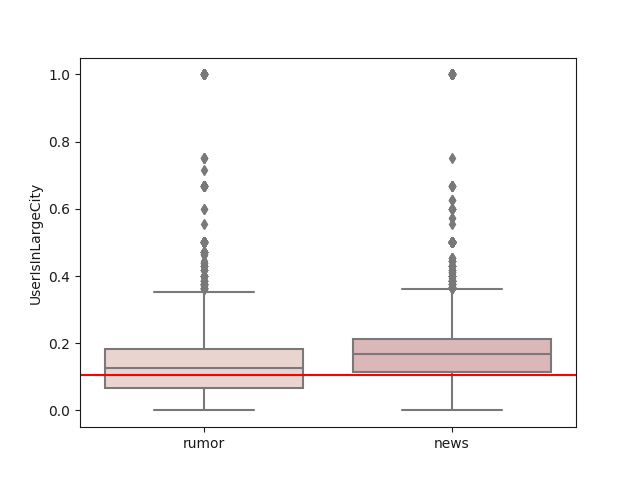}
}
\caption{Creditscore and some strong features for \textit{Munich shooting} in red lines, compared with the corresponding average scores for \textit{rumor} and \textit{news}.}
\end{figure}

\section{Conclusion}
\label{sec:conclusion}


In this work, we present a deep analysis on the feature variants over 48 hours for the rumor detection task. The results show that the low-level hidden representation of tweets feature is at least the second best features over time. We also derive explanations on the low performance of \textit{supposed-to-be-strong} high-level features at early stage. The study also indicates that, there is still considerable room to improve the effectiveness of the neural network-based rumor detection methods, e.g., by leveragining the embeddings from different sources rather than only text contents.
\bibliographystyle{abbrv}
\vspace{-0.2cm}

\bibliography{websci17}

\end{document}